\documentclass{aa}  

\usepackage{natbib}
\usepackage{graphicx}
\usepackage{txfonts}
\usepackage{hyperref}
\usepackage{url}

\begin{document} 
  \title{ODUSSEAS: A machine learning tool to derive effective temperature and metallicity for M dwarf stars}
  
  \author{A. Antoniadis-Karnavas 
          \inst{1,2}, S. G. Sousa\inst{1}, E. Delgado-Mena\inst{1}, N. C. Santos\inst{1,2}, G. D. C. Teixeira\inst{1,2}, V. Neves\inst{3,4} }
          
   \institute{Instituto de Astrofísica e Ciências do Espaço, Universidade do Porto, CAUP, Rua das Estrelas, 4150-762 Porto, Portugal.\\
  \email{alexandros.antoniadis@astro.up.pt}
 \and
 Departamento de Física e Astronomia, Faculdade de Ciências, Universidade do Porto, Rua do Campo Alegre, 4169-007 Porto, Portugal. 
 \and
 Instituto Federal do Paraná, 85860000, Campus Foz do Iguaçu, Foz do
Iguaçu-PR, Brazil.
 \and{Casimiro Montenegro Filho Astronomy Center, Itaipu Technological Park,
85867-900, Foz do Iguaçu-PR, Brazil. \\ }}

  \date{Received / Accepted}

  \abstract
  {}
   {The derivation of spectroscopic parameters for M dwarf stars is very important in the fields of stellar and exoplanet characterization. The goal of this work is the creation of an automatic computational tool, able to derive quickly and reliably the T$_{\mathrm{eff}}$ and [Fe/H] of M dwarfs by using their optical spectra, that can be obtained by different spectrographs with different resolutions.}
   {ODUSSEAS (Observing Dwarfs Using Stellar Spectroscopic Energy-Absorption Shapes) is based on the measurement of the pseudo equivalent widths for more than 4000 stellar absorption lines and on the use of the machine learning Python package "scikit-learn" for predicting the stellar parameters.}
   {We show that our tool is able to derive parameters accurately and with high precision, having precision errors of $\sim$30 K for T$_{\mathrm{eff}}$ and $\sim$0.04 dex for [Fe/H]. The results are consistent for spectra with resolutions between 48000 and 115000 and SNR above 20. }
   {}
   {}

\titlerunning{Machine learning tool ODUSSEAS for M dwarf stars}
\authorrunning{A. Antoniadis-Karnavas et al.}

\maketitle

\section{Introduction}

Spectra can be used to reveal the chemical composition of the stars, as well as important stellar atmospheric parameters, such as effective temperature (T$_{\mathrm{eff}}$) and [Fe/H]. These parameters are crucial for the characterization of the stars and therefore fundamental to understand their formation and evolution. Furthermore, they influence the properties of the planets forming and orbiting around them \citep{ever13}. However, the spectroscopic analysis to derive these parameters has some difficulties to overcome. One of the main problems is the correct determination of the spectral continuum, which is more problematic in cool and faint stars, such as M dwarfs. Their study is quite difficult and complicated, compared to FGK stars, since in M dwarfs, molecules are the dominant sources of opacity. These molecules create thousands of lines that are poorly known and moreover many of them blend with each other. Therefore, the position of the continuum is hardly identified in their spectra. 
 
Methods which rely on the correct determination of the continuum, work better only for the metal poor and earliest types of M dwarfs \citep{woolf05}. Methods using spectral synthesis have not achieved as precise results as in FGK cases, because of the poor knowledge of many molecular line strengths.  
Recently, spectral synthesis in the near infrared has presented advances, as shown by several studies. \citep{one12,lind16,raj18, pass19}.
  
Regarding these limitations, most  attempts for determining effective temperature and metallicity, are done with photometric calibrations \citep{bon05,johnapps09,neves12} or spectroscopic indices \citep{rojas10,rojas12,mann13a}. Metallicity uncertainties range from 0.20 dex using photometric calibrations, to 0.10 dex by using spectroscopic scales in the infrared \citep{rojas12}. For T$_{\mathrm{eff}}$, precisions of 100 K are reported, but significant uncertainties and systematics are still present, ranging from 150 to 300 K. \citep{casag08,rojas12}.

One of the most popular methods to derive atmospheric stellar parameters for FGK stars is by measuring the equivalent widths (EW) of many metal lines of the spectrum.  
\citet{neves14} using the MCAL code, measured pseudo EWs in the optical part of the spectrum for 110 M dwarfs observed in the HARPS GTO M dwarf program, by setting a pseudo continuum for each line. They proceeded to the derivation of T$_{\mathrm{eff}}$ and [Fe/H] of these stars applying a calibration based on reference photometric T$_{\mathrm{eff}}$ and [Fe/H] scales that exist for 65 of them from \citet{casag08} and \citet{neves12} respectively. In the first case, the reference T$_{\mathrm{eff}}$ is the average value of the V - J, V - H, and V - K photometric scales as seen in \citet{casag08}, while for [Fe/H] the calculation of its reference values was done using stellar parallaxes, V and Ks magnitudes as described in \citet{neves12}.

Machine learning is an increasingly popular concept  in several fields of science. It can be accurate in predicting outcomes without the need of the user explicitly creating a specific model to the problem at hand. The algorithms in machine learning receive input data and by applying statistical analysis, they predict an output value within a reasonable range. 
The interest for machine learning algorithms and automatic processes in astronomy is emerging from the increasing volume of survey data \citep{how17}. It can be applied to a wide range of studies, with the input attributes being for example the photometric properties of the sources \citep{dasa19,akras19,rau19,ucci19}. 

In our work, we follow the pseudo-EW approach. 
We present our tool ODUSSEAS (Observing Dwarfs Using Stellar Spectroscopic Energy-Absorption Shapes), which makes use of the machine learning "scikit learn" package of Python. It offers a quick automatic derivation of T$_{\mathrm{eff}}$ and [Fe/H] for M dwarf stars, by being provided with their 1D spectra and their resolutions. 
The main advantage of this tool, compared to other ones that derive stellar parameters such as the MCAL code by \citet{neves14} (which is limited to HARPS range and needs manual adjustment of results for different resolutions), is that it can operate simultaneously in an automatic fashion for spectra of different resolutions and different wavelength ranges in the optical. It is based on a supervised machine learning algorithm, meaning that it is provided with both input and expected output for creating a model. 
This input to the machine learning function are the values of the pseudo EWs for 65 HARPS spectra and the expected output are the values of their reference T$_{\mathrm{eff}}$ and [Fe/H] from \citet{casag08} and \citet{neves12} respectively. 
After training with a part of these HARPS data, the algorithm produces a model and tests it on the rest of the HARPS data. 
It predicts their values and compares them with the reference ones given as expected output. Thus, it examines the accuracy and the precision of the model by using several regression metrics described later. Finally, it applies the model to unknown spectra and estimates their stellar parameters. 

In Sect.~\ref{EWsection} we describe how the tool computes the pseudo EWs. 
In Sect.~\ref{mlsection} we describe our tool and the flow of its process. We explain the characteristics of the machine learning function and its efficiency regarding different regression types, resolutions and wavelength areas.
In Sect.~\ref{otherspec} we apply our tool to spectra obtained by several spectrographs of various resolutions and we examine the results. 
Finally, Sect.~\ref{sum} summarizes the work presented in this paper.


\section{Pseudo-EW measurements}
\label{EWsection}
  
Since the identification of the continuum is very difficult in the spectra of M dwarfs, we follow the way of setting a pseudo continuum in each absorption line. The method is based on measurements of the pseudo EWs of absorption lines and blended lines in the range between 530 and 690 nm. We have excluded the parts where the activity-sensitive Na doublet and H$\alpha$ lines and strong telluric lines reside. The linelist consists of 4104 features. It is given in the form of left and right boundaries, between which these absorption features are supposed to be created. This method, based on pseudo EWs and the specific linelist, was used by \citet{neves14}.  

We have created our own Python version of the method to compute the pseudo EWs.
Our code reads the linelist and the 1D fits files of the stellar spectra. We have set an option for radial-velocity correction of the input spectra by our code, in the case they are shifted. Then, for each line, it identifies the position of the minimum flux of the feature, which is the central absorption wavelength. Starting from it, the code identifies the maximum in each side of this absorption feature, after having cut this spectral area at the range defined by the respective boundaries provided in the linelist. Eventually, it fits the pseudo continuum along the edges of the absorption feature with a straight line and it obtains the pseudo EW by calculating the area between the pseudo continuum and the flux.  
Mathematically, the pseudo EW is defined as following, where F$_{pp}$ is the value of the flux between the peaks of the feature (i.e. the pseudo continuum) and F$_{\lambda}$ is the flux of the line at each integration step.

\begin{equation}
{\mathrm{pseudoEW}} = \Sigma \frac{(F_{pp}-F_{\lambda})}{F_{pp}} \Delta{\lambda}
\end{equation}

\begin{figure}
\centering
\includegraphics[width= \hsize]{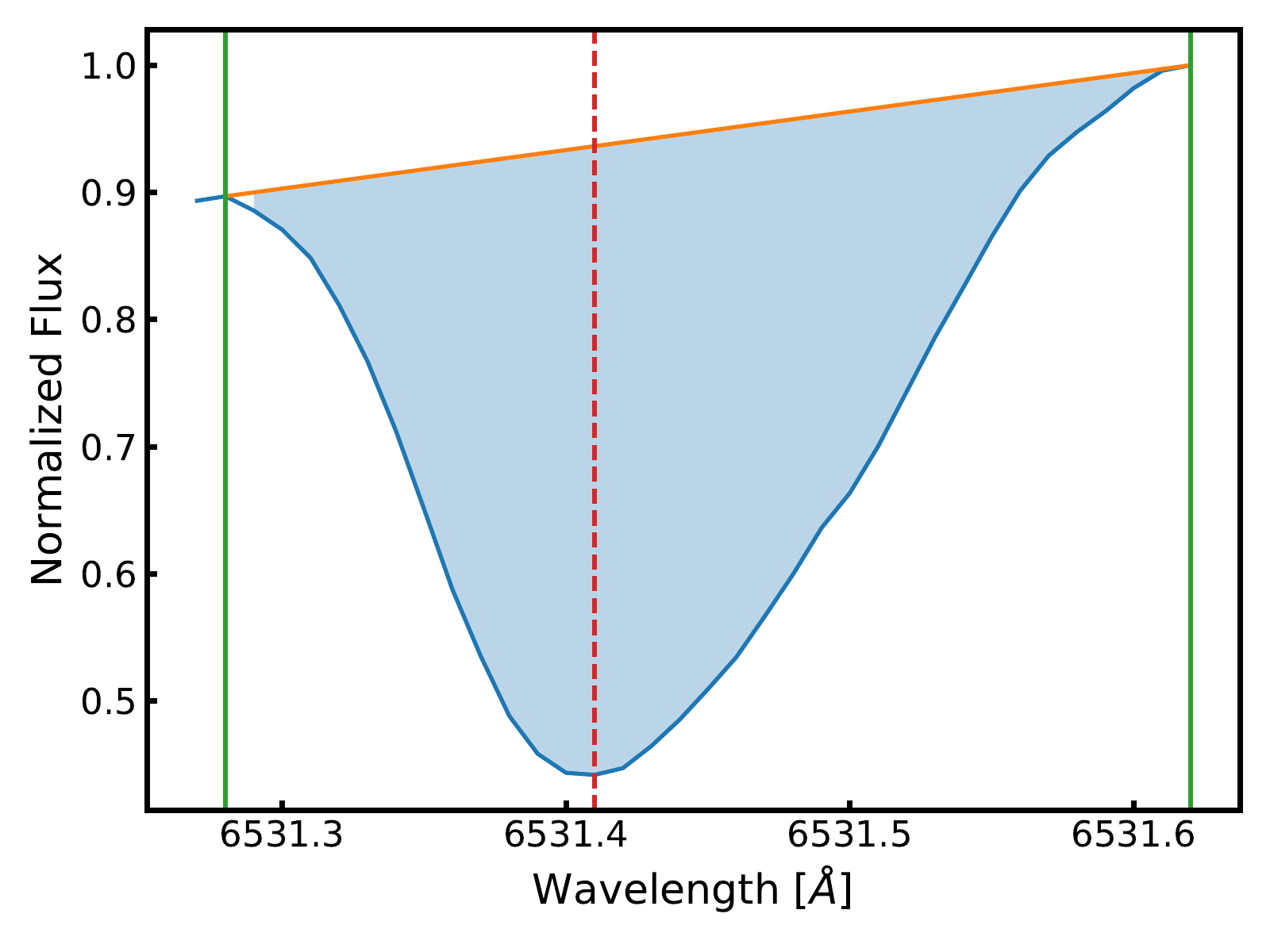} \\
\caption{Area fitting for the calculation of pseudo EW for a line with central $\lambda$ = 6531.4 $\AA$  of the star Gl176. The position of the pseudo continuum is adjusted accordingly. This pseudo EW is equal to 87 m$\AA$. }
\label{EWs} 
\end{figure}

We present such example in Fig.~\ref{EWs} where we use the star Gl176 and an absorption line at the region around 6530 $\AA$.

The evaluation of our pseudo-EW measurements, by comparing them with the ones obtained from MCAL code, is presented at Appendix~\ref{A}.

\section{Machine learning on M dwarfs}
\label{mlsection}

We base our tool for the derivation of T$_{\mathrm{eff}}$ and [Fe/H] on the machine learning concept.  
The user needs to run two codes. The "HARPS$\_$dataset.py" creates the databases which contain pseudo-EW measurements in different resolutions and the reference stellar parameters. The "ODUSSEAS.py" measures the pseudo EWs of new stellar spectra and derives their unknown T$_{\mathrm{eff}}$ and [Fe/H] via machine learning.
Below, we explain the details of their structure, describing the input parameters and how to use the codes.

\subsection{The HARPS dataset}    

\begin{figure}
\centering
\includegraphics[width= \hsize]{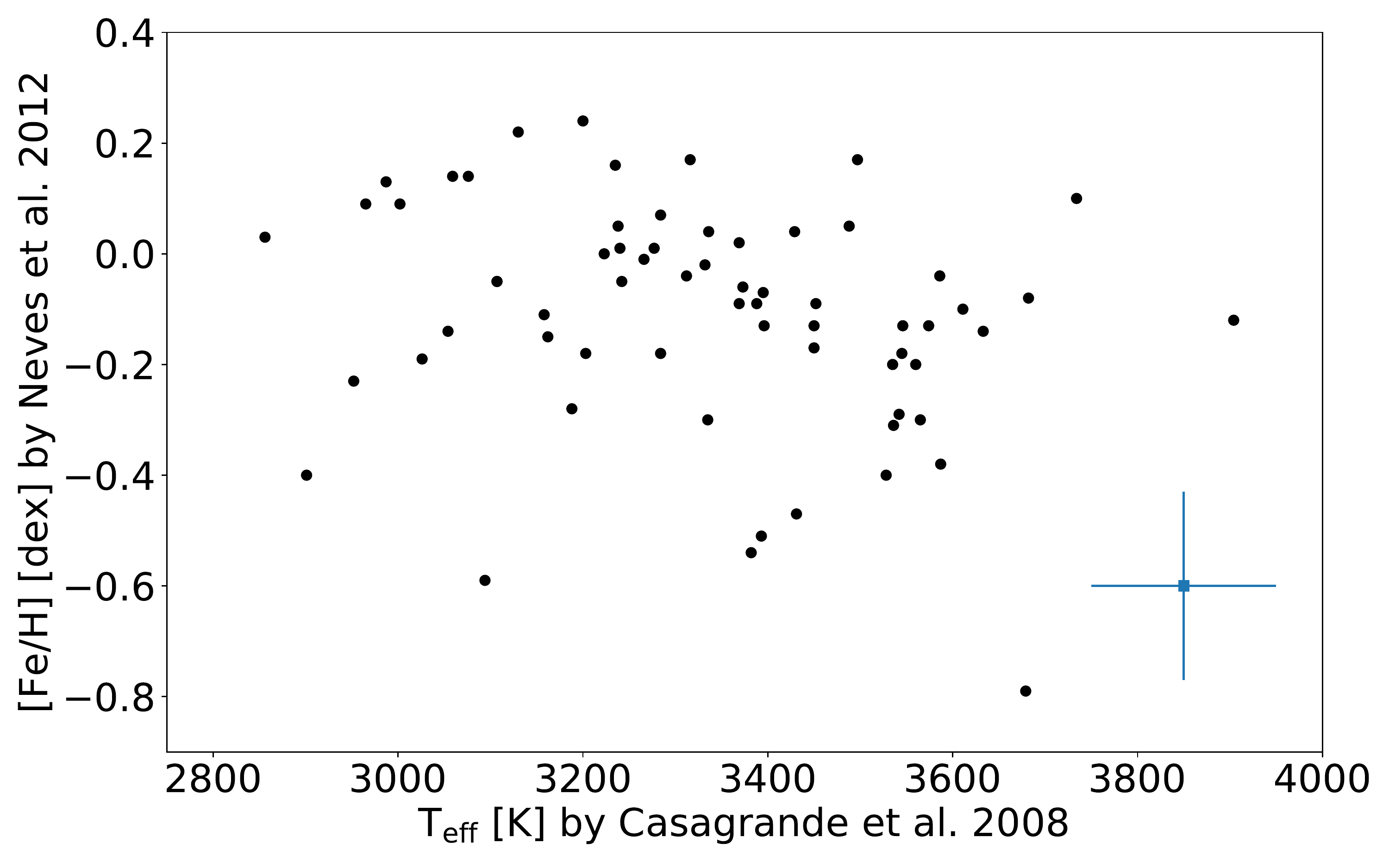} \\
\caption{The distribution of reference T$_{\mathrm{eff}}$ and [Fe/H] of the 65 stars used to train and test the machine learning models. The cross represents the uncertainties of their photometric derivations, which are 100 K and 0.17 dex respectively.}
\label{range} 
\end{figure}

Each time the code "HARPS$\_$dataset.py" runs, the outcome is a file which is used later as input to the machine learning algorithm when running "ODUSSEAS.py" for training the machine and testing the generated model. It contains the names of 65 stars of the HARPS M dwarf sample, the central wavelengths of the 4104 absorption features from 530 to 690 nm, their pseudo-EW values according to the resolution we convolve the spectra and their reference values of T$_{\mathrm{eff}}$ and [Fe/H]  from \citet{casag08} and \citet{neves12} respectively. All of these 65 spectra have SNR above 100, as reported by \citet{neves14}.  
They are presented in Table~\ref{refparam}. The range of the reference stellar parameters is presented in Fig~\ref{range}. Their photometric derivations have uncertainties of 100 K for T$_{\mathrm{eff}}$ and 0.17 dex for [Fe/H], as reported by \citet{casag08} and \citet{neves12} respectively.

The convolution function we use is the "instrBroadGaussFast" of "pyAstronomy" (\url{https://github.com/sczesla/PyAstronomy}), which applies Gaussian instrumental broadening. The width of the kernel is determined by the resolution. 
A description of it can be found at \url{https://www.hs.uni-hamburg.de/DE/Ins/Per/Czesla/PyA/PyA/pyaslDoc/aslDoc/broad.html}.

Since the HARPS spectra have a specific finite resolution, our code calculates the actual resolution to which they need to be convolved by the function, in order to get spectra to the final resolution we really want. 
This calculation is done considering the following relation: 
\begin{equation}
\sigma _{conv} = \sqrt{\sigma _{final}^{2} - \sigma _{orig}^{2}}
\end{equation}

where $\sigma_{conv}$ corresponds to the resolution to which we need to convolve a spectrum with original resolution of $\sigma_{orig}$, in order to get a final resolution of $\sigma_{final}$.

The settings input by the user are two. a) Choose whether or not to convolve the reference HARPS spectra to the spectral resolution of our new data. We already provide precomputed pseudo EWs for a range of spectral resolutions in widely used spectrographs.
In that case there is no need to convolve again the spectra and recalculate the pseudo EWs.
b) The resolution of the data we want to analyse. 

\begin{figure*}
\centering
$\begin{array}{c}
\includegraphics[width=11cm]{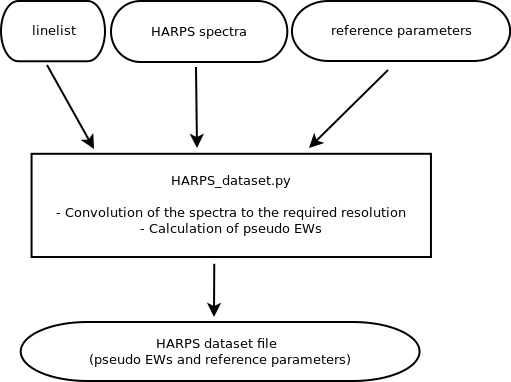} \\
\end{array}$
\caption{ The workflow of HARPS$\_$dataset.py}
\label{HARPSdia} 
\end{figure*}

The "HARPS$\_$dataset.py" is presented schematically in Fig.~\ref{HARPSdia}.

\subsection{ODUSSEAS tool}

"ODUSSEAS.py" makes use of two algorithms that we developed: the "New$\_$data.py", for measuring the pseudo EWs of new spectra to analyze, and the "MachineLearning.py" for the derivation of their T$_{\mathrm{eff}}$ and [Fe/H]. The innovative aspect of this tool is the simultaneous predictions for spectra of different resolutions and wavelength ranges.

The user has the option to activate the automatic radial velocity correction for the spectra if they are shifted.
In addition, the user can set the regression type to be used by the machine learning process. The "ridge" is recommended, but also "ridgeCV" and "linear" work at similar level of efficiency as well. We present the efficiency of all the regression types used in Sect.~\ref{mleff}.

The workflow of "New$\_$data.py" is similar to the "HARPS$\_$dataset.py". 
It reads the files and resolutions of new spectra and, if needed, it calculates and corrects their radial velocity shift.
In addition, if the original step of a spectrum is not 0.010, i.e. equal to that of the HARPS dataset, it is changed with linear interpolation to this value. Thus, the pseudo EWs are measured in a consistent way.
The files containing the pseudo-EW measurements of each spectrum are then used during the operation of "MachineLearning.py", which returns the values of T$_{\mathrm{eff}}$ and [Fe/H] along with the regression metrics of the models that predicted them.

\begin{figure*}
\centering
$\begin{array}{c}
\includegraphics[width= 11cm ]{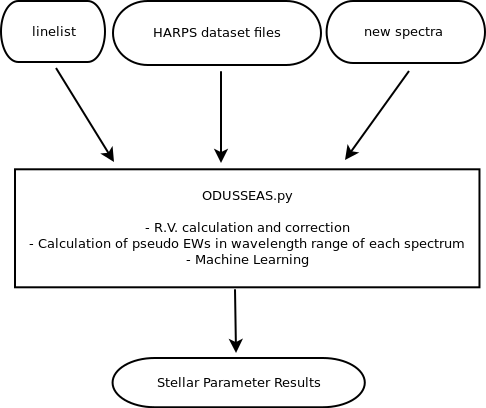} \\
\end{array}$
\caption{ The workflow of ODUSSEAS.py}
\label{Mdwarfsdia} 
\end{figure*}

The diagram of "ODUSSEAS.py" is presented in Fig.~\ref{Mdwarfsdia} showing concisely its inputs, operations and output.

\subsection{Machine learning function}
\label{mlf}

Here we present in more detail the machine learning function.
The machine learning algorithm operates in a loop for each star separately, as each star may have different wavelength range and different resolution. For each star in the filelist, it loads automatically two files: the HARPS dataset of respective resolution, for training and testing the model, and the pseudo EWs of the star for which we want to derive T$_{\mathrm{eff}}$ and [Fe/H], in order to apply the model and return the stellar parameters. Based on the wavelength range that each spectrum has, a mask is applied on the HARPS dataset for considering the absorption lines in common. 

The 65 HARPS stars split into training group consisting of the 70\% of the sample (45 stars)  and into testing group consisting of the remaining 30\% of the population (20 stars). With these numbers selected, the machine learning model can be both trained accurately and tested on a sufficient number of stars.

We provide the algorithm with different regression types that can be used: the "linear", the "ridge", the "ridgeCV", the ''multi-task Lasso" and the "multi-task Elastic Net". 
All these kinds of models provide an output value by fitting a linear regression to the input values.
The relation between the predicted value \textit{y} (the stellar parameter), the input variables \textit{x} (the pseudo EWs) and the coefficients \textit{w} is  expressed as 
\begin{equation}
y(w,x)=w_{o}+w_{1}x_{1}+...+w_{p}x_{p}
\end{equation}

The mathematical details of each regression type are described in the official online documentation at \url{https://scikit-learn.org/stable/modules/linear_model.html}.
 
The performance of  machine learning is indicated by the following three kinds of regression metrics that are returned. 
The mean absolute error is computed when the model is applied on the test dataset. It corresponds to the expected value of the absolute error loss in the predictions.
In addition, the "explained variance score" is calculated. The best possible value of this score is 1.0. Variance is the expectation of the squared deviation of a random variable from its mean. It measures how far a set of numbers are spread out from their average value.
Furthermore, the "r2 score" computes the coefficient of determination, defined as R$^2$. The coefficient of determination is the proportion of the variance in the dependent variable that is predictable from the independent variables. This score provides a measure of how well future samples are likely to be predicted by the model. Best possible score is 1.0 too. A constant model that always predicts the expected value, disregarding the input features, would get a score of 0.0.
In our case of multi-output, the resulting "explained variance" and "r2" scores are by default the averages with uniform weight of the respective scores for T$_{\mathrm{eff}}$ and [Fe/H].  
The mathematical types of those regression metrics are described in their official online address at \url{https://scikit-learn.org/stable/modules/model_evaluation.html#regression-metrics}.

\begin{figure}
\centering
$\begin{array}{c}
\includegraphics[width=\hsize]{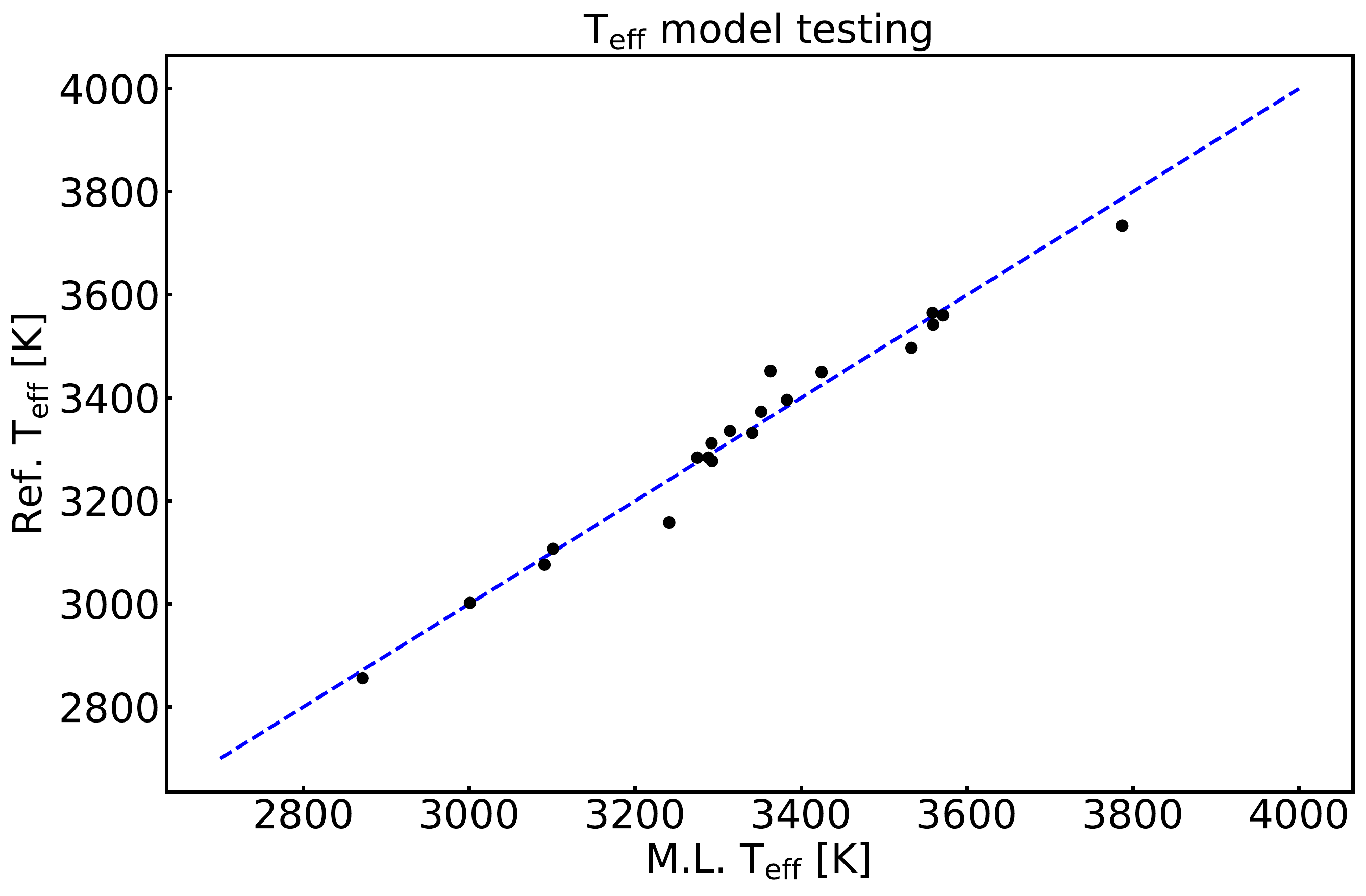}  \\
\includegraphics[width=\hsize]{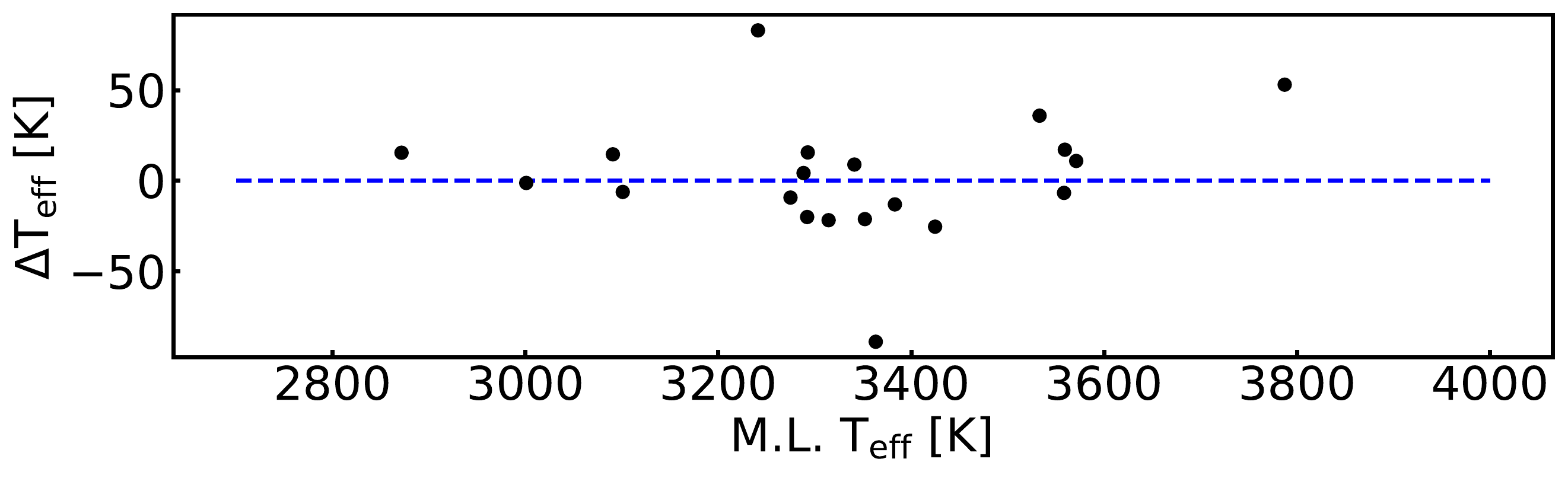}  \\
\\
\includegraphics[width=\hsize]{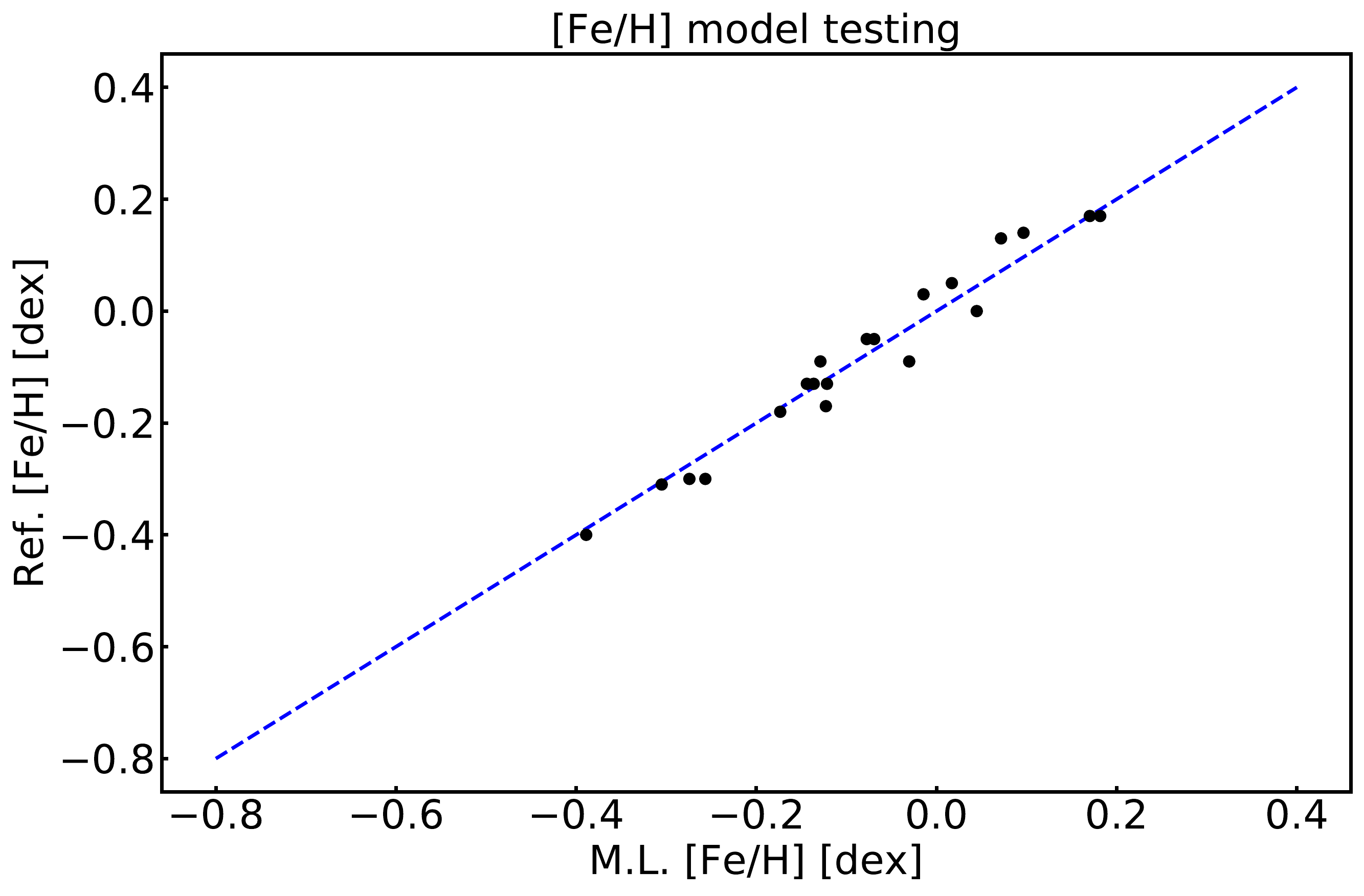}  \\
\includegraphics[width=\hsize]{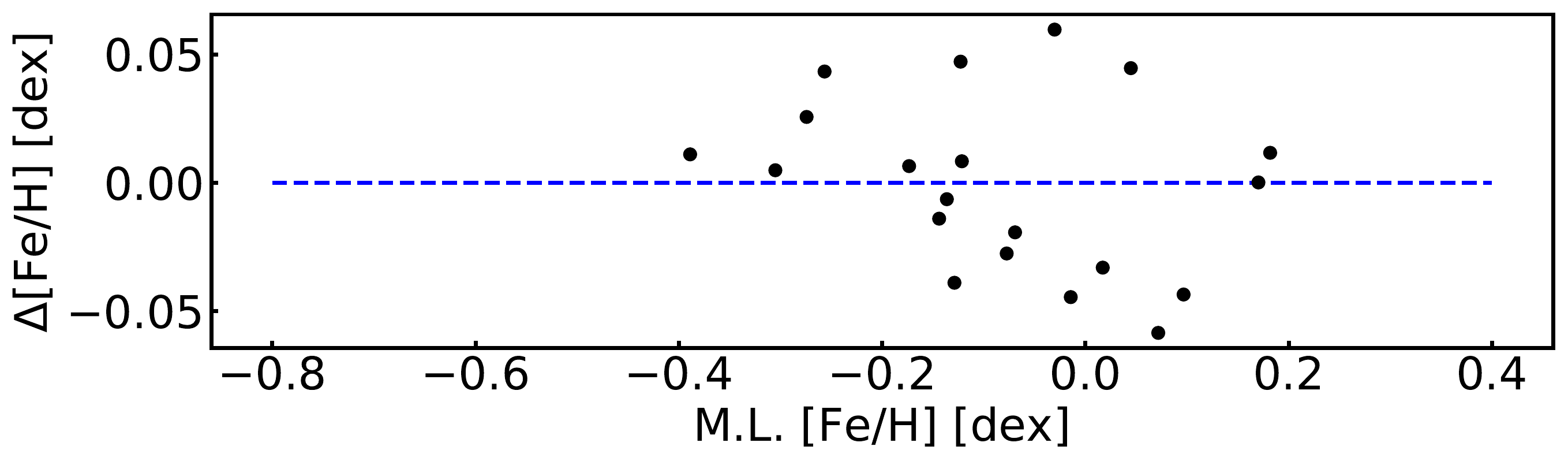}  \\
\end{array}$
\caption{Demonstration of predictions applying ridge regression. Upper panel: the T$_{\mathrm{eff}}$ values expected (Ref.) and predicted (M.L.) on the test dataset, along with their differences. Lower panel: the [Fe/H] values expected (Ref.) and predicted (M.L.) on the test dataset, along with their differences.}
\label{ml} 
\end{figure}

For each star, the tool makes 100 determinations by splitting randomly the train and test groups each time. After these determinations, it returns the average values of T$_{\mathrm{eff}}$ and [Fe/H], the average values of the mean absolute errors of the models, the average scores of machine learning and the dispersion of T$_{\mathrm{eff}}$ and [Fe/H] (measured as the standard deviation).
This iterative process minimizes the possible dependence of the resulting parameters on how the stars from the HARPS dataset are split for training and testing in one single measurement. 
Since the reference stars are only 65, which stars end up in the training set could change the results in a measurement. This is the reason we do these multiple runs with shuffling and splitting the reference stars in different train and test groups, and finally we calculate the average values and the dispersion.
The final results are automatically saved in the file called "Parameter$\_$Results.dat". 
Moreover, it saves a group of plots with the reference and the predicted parameters of model testing, as well as their differences, as a visualization of the model accuracy. An example is presented at Fig.~\ref{ml}.

\subsection{Machine learning efficiency}
\label{mleff}

Firstly, we test the regression models mentioned above to find the best one. We use the original spectra of the HARPS dataset to their real resolution of 115000. For 100 runs with each regression type, we measure the scores and the absolute mean errors of the stellar parameters on the test set. We report the average values around which each model tends to result in Table~\ref{reg}. 
The "linear", "ridge" and "ridgeCV" work very well in general, having "r2" and "explained variance" scores with average values around 0.93 and 0.94 respectively. The range of these scores, in the 100 runs, is usually from 0.87 to 0.99. 
The average uncertainties of those regression types are $\sim$27 K for T$_{\mathrm{eff}}$ and $\sim$0.04 for [Fe/H]. The "ridge" model has slightly greater scores than the "linear" one.  
"RidgeCV", which has a built-in cross validation function that applies "leave-one-out" or "k-fold" strategies, does not seem to work better than the classic "ridge" one, at least in this sample of M dwarf measurements. Furthermore, "multi-task Elastic Net" and "multi-task Lasso" give considerably lower scores and higher mean absolute errors. Thus, we suggest "ridge" regression, as it operates best on the spectral values of the M dwarfs.

Secondly, we evaluate the "explained variance" and "r2" scores and the mean absolute errors of the algorithm for different resolutions of the spectra. We do it for the HARPS dataset at its actual resolution of 115000 and we repeat this test for convolved datasets at resolutions of other broadly used spectrographs: 110000 (UVES),  94600 (CARMENES) , 75000 (SOPHIE) and 48000 (FEROS). 
This is done to examine the level of machine learning precision towards lower resolutions.
After 100 measurements of each case, we present the average values at Table~\ref{reso}.

To further test the reliability of the method, we examine the efficiency of the machine learning in different wavelength ranges of the spectrum. We divide the linelist, which is from 530 to 690 nm, in four spectral regions and we calculate the respective scores and mean absolute errors. We do this test to check if machine learning works better using the full range or a specific part of the wavelengths. 
For this test, we use the case of the convolved data at the resolution of 110000. 
The machine learning operates at its best while using the full range of the initial linelist. In addition, regarding the divided areas, we notice that the bluer the part the higher the scores and the lower the mean absolute errors respectively. In general, the results show that we can get highly precise predictions for stars observed at any part of the 530-to-690 nm spectrum. These results are presented in Table~\ref{linelist}. 

\begin{table*}
\centering
\caption{The average values of the scores and the mean absolute errors (M.A.E.) for T$_{\mathrm{eff}}$ and [Fe/H] of the test dataset, after 100 runs of each regression type.}
\begin{tabular}{ccccc}
\hline\hline\\
Regression & r2 score & E.V. score & M.A.E. T$_{\mathrm{eff}}$ & M.A.E. [Fe/H]  \\
 & & & [K] & [dex] \\
\hline\\
Ridge & 0.93 & 0.94 & 27 & 0.037 \\
RidgeCV & 0.93 & 0.94 & 27 & 0.038 \\
Linear & 0.93 & 0.93 & 27 & 0.039 \\
Multi-task Elastic Net & 0.91 & 0.92 & 35 & 0.045 \\
Multi-task Lasso & 0.88 & 0.89 & 41 & 0.056 \\
\hline\\
\end{tabular}
\label{reg}
\end{table*}

\begin{table*}
\centering
\caption{The average values of the scores and the mean absolute errors (M.A.E.) for T$_{\mathrm{eff}}$ and [Fe/H] of the test dataset, after 100 runs of each resolution (using the "ridge" regression).}
\begin{tabular}{ccccc}
\hline\hline\\
Resolution & r2 score & E.V. score & M.A.E. T$_{\mathrm{eff}}$ & M.A.E. [Fe/H]  \\
 & & & [K] & [dex] \\
\hline\\
real 115000 & 0.93 & 0.94 & 27 & 0.037 \\
conv. 110000 & 0.93 & 0.94 & 28 & 0.038 \\
conv. 94600 & 0.93 & 0.93 & 28 & 0.039 \\
conv. 75000 & 0.93 & 0.93 & 29 & 0.041 \\
conv. 48000 & 0.92 & 0.93 & 30 & 0.043 \\
\hline\\
\end{tabular}
\label{reso}
\end{table*}

\begin{table*}
\centering
\caption{The average values of the scores and the mean absolute errors (M.A.E.) for T$_{\mathrm{eff}}$ and [Fe/H] of the convolved-to-110000 dataset, after 100 runs of each wavelength part of the linelist.}
\begin{tabular}{cccccc}
\hline\hline\\
Wavelength range & Number of lines & r2 score & E.V. score & M.A.E. T$_{\mathrm{eff}}$ & M.A.E. [Fe/H]  \\
(nm) & & & & [K] & [dex] \\
\hline\\
530 - 690 & 4104 & 0.93  & 0.94 & 28 & 0.038 \\
530 - 580 & 1300 & 0.92 & 0.93 & 31 & 0.039 \\
580 - 630 & 1300 & 0.91 & 0.91 & 48 & 0.044 \\
630 - 690 & 1504 & 0.89 & 0.90 & 56 & 0.048 \\
\hline\\
\end{tabular}
\label{linelist}
\end{table*}

\section{Derivation of stellar parameters}
\label{otherspec}

We apply our tool to spectra obtained by five  widely used instruments of different resolutions: HARPS of 115000, UVES of 110000, CARMENES of 94600, SOPHIE of 75000 and FEROS of 48000.
The spectra were taken from the respective public data archives.
To test the efficiency of our tool on other-than-HARPS instruments, we use spectra from stars in common with the HARPS dataset, so we can compare their results with the reference parameters of the respective HARPS spectra.
To validate further the accuracy of our tool, we proceed to determinations and comparisons on more stars. Finally, we discuss about possible future improvements of our determinations.

\subsection{Resolution and spectral shape}
\label{s1}

\begin{figure}
\centering
$\begin{array}{c}
\includegraphics[width=\hsize]{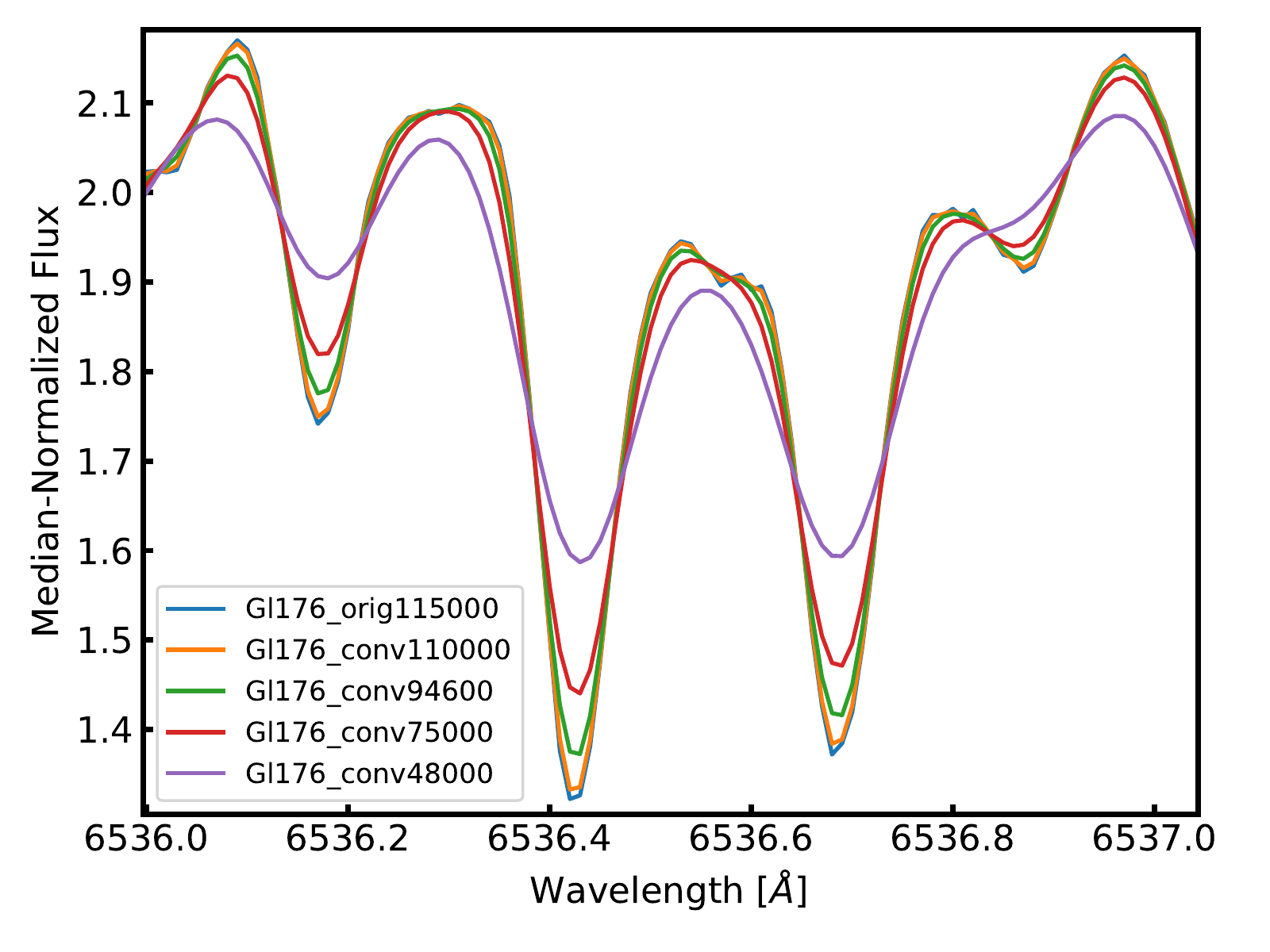}  \\
\end{array}$
\caption{ The shape of the HARPS original spectrum for Gl176 and convolved in different resolutions. The lower the resolution the swallower the absorption lines.}
\label{convall} 
\end{figure}

\begin{figure}
\centering
$\begin{array}{c}
\includegraphics[width=\hsize]{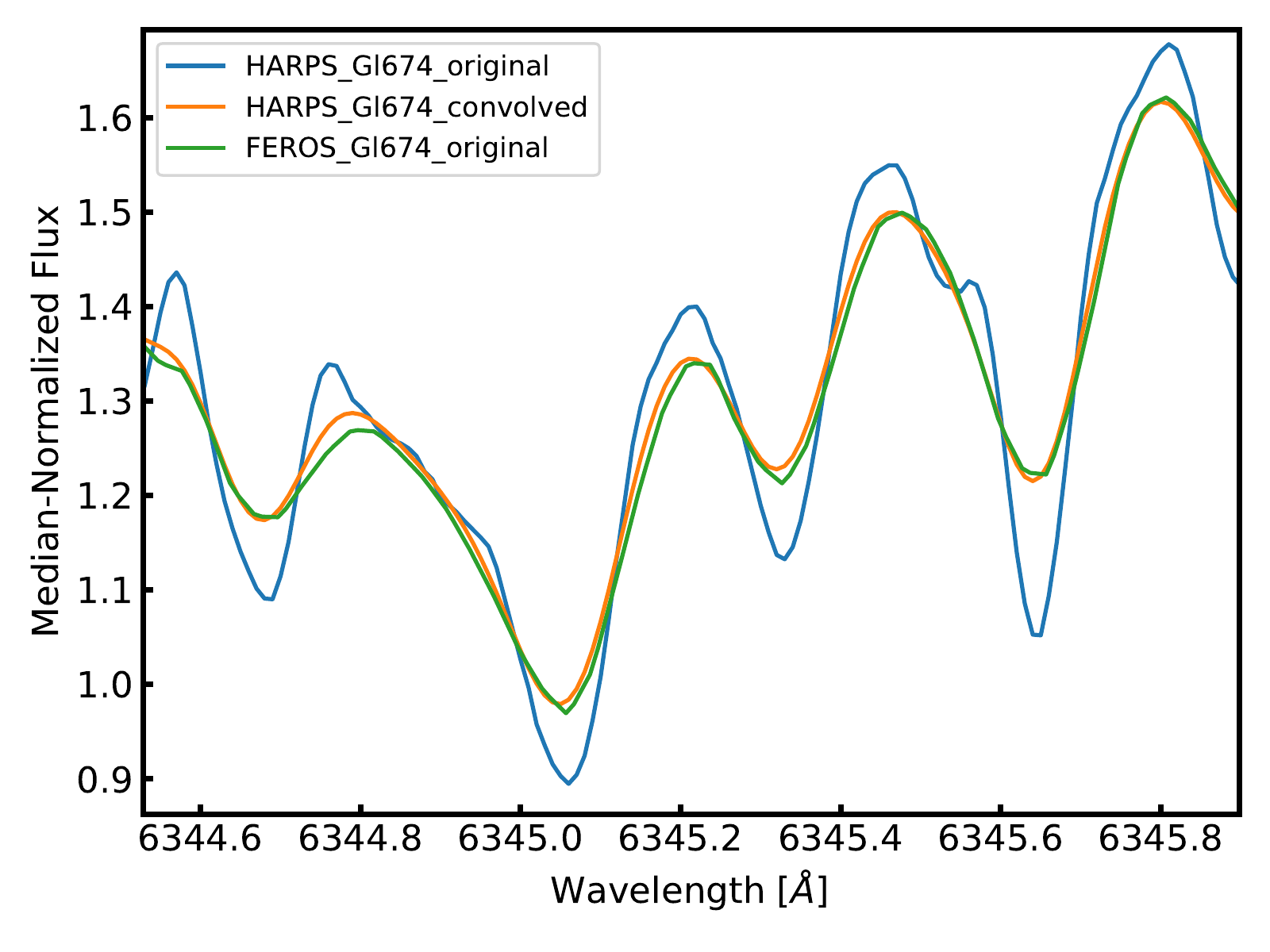}  \\
\end{array}$
\caption{ The shape of spectra for Gl674 in original HARPS resolution (blue), FEROS resolution (green) and HARPS convolved to FEROS resolution (orange).}
\label{shapeferos} 
\end{figure}

\begin{figure}
\centering
$\begin{array}{c}
\includegraphics[width=\hsize]{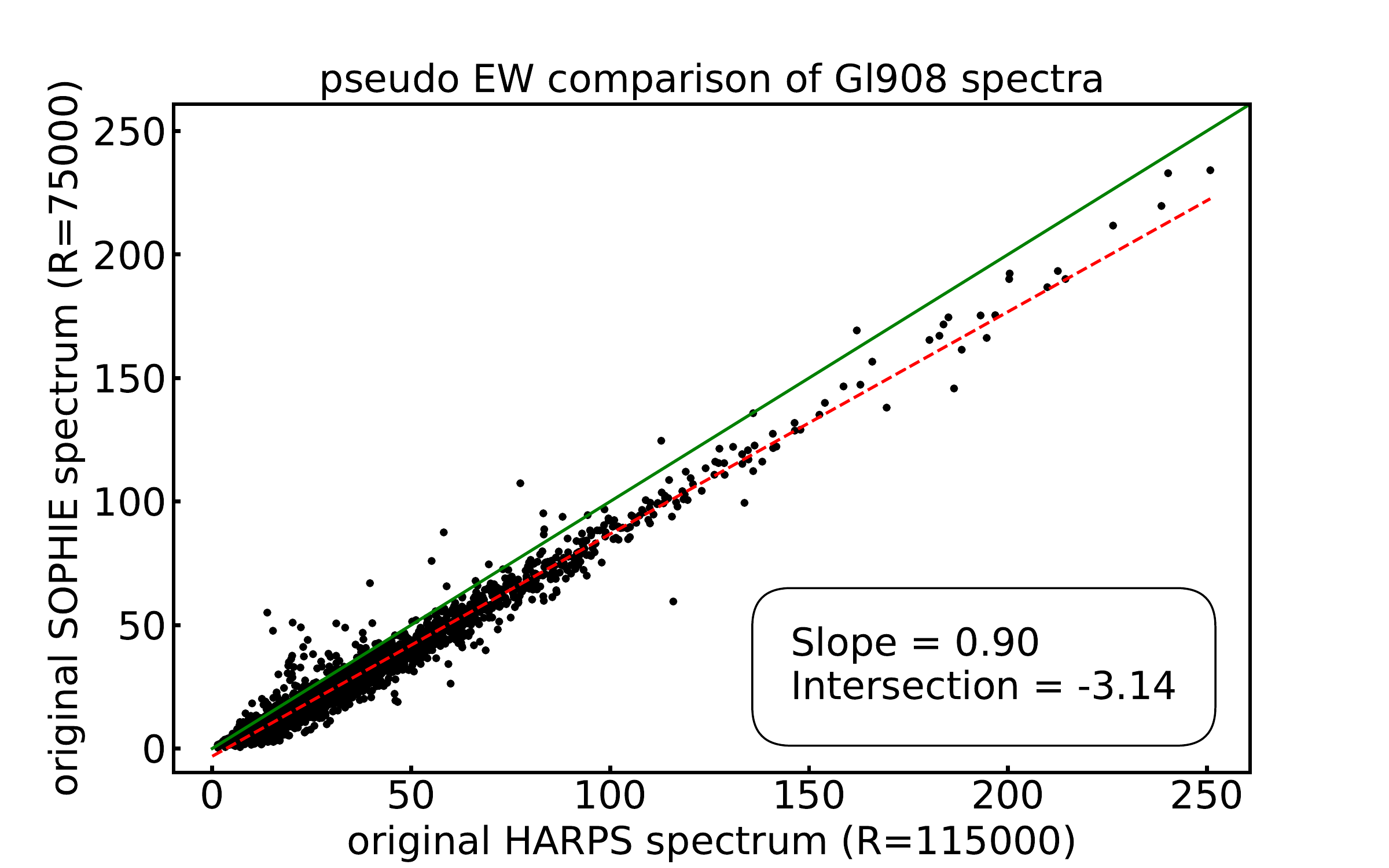}  \\
\includegraphics[width=\hsize]{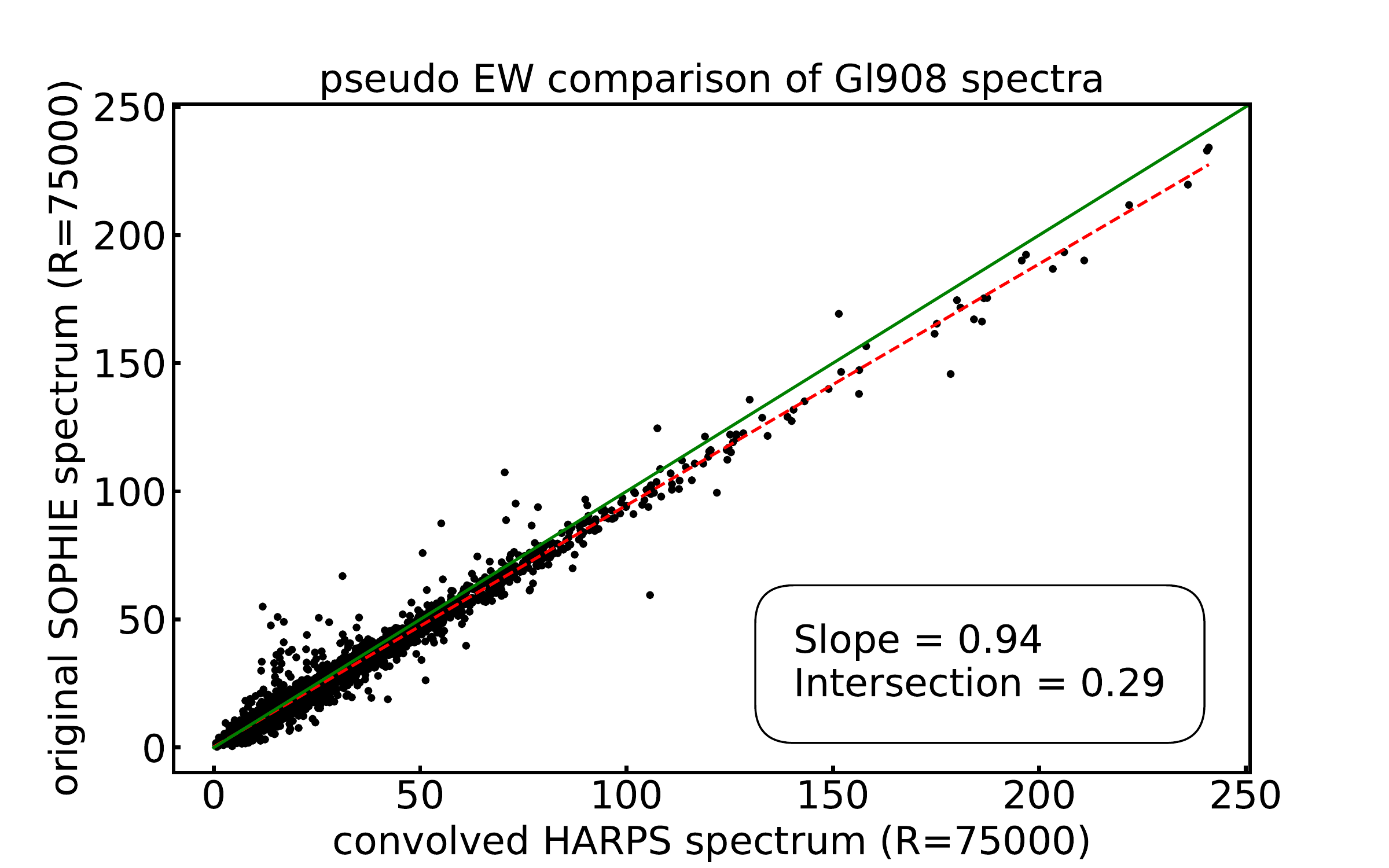}  \\
\end{array}$
\caption{ Upper panel: pseudo-EW values of the original Gl908 spectra for HARPS and SOPHIE. Lower panel: pseudo-EW values of Gl908 after the convolution of HARPS to the resolution of SOPHIE. The units of pseudo EWs are m$\AA$. After the convolution, there is agreement between the identity line (solid green) and the slope (dashed red), with the intersection being close to 0.}
\label{hsEW} 
\end{figure}

We examine the spectral change of M dwarfs according to convolution in different resolutions.
The shapes of M dwarf spectra are different when obtained in lower resolutions.
In general, the lower the resolution, the shallower the absorption lines. 
This is illustrated in Fig~\ref{convall} where three lines of Gl176 are shown in detail, for the original HARPS spectrum and the convolved ones to several resolutions. 
We also measure these lines and we report their pseudo-EW values in Table~\ref{EWvalues}, to show their differences. 
The relative differences can vary, as not only the depth changes but also the location of the pseudo continuum is different in each case. They all confirm that the lower resolution always has lower pseudo-EW values.
This is why we need to convolve the HARPS spectra to the respective resolutions of the new spectra. Consequently, machine learning compares the pseudo EWs of the same resolution and predicts accurately the stellar parameters.
In Fig~\ref{shapeferos} we show the spectral shapes of Gl674 for three different cases: the original HARPS spectrum with resolution 115000, the convolved HARPS spectrum to the resolution of FEROS (48000) and the original FEROS spectrum that is the lowest resolution we examine. We notice that the convolved HARPS spectrum follows the shape of the FEROS one in a consistent way.
In Fig~\ref{hsEW}, we show the comparison of the pseudo EWs of  SOPHIE spectrum for Gl908 and the spectrum of the same star by HARPS before and after its convolution. The SOPHIE spectrum, which is of lower resolution, has consistently lower pseudo-EW values than the HARPS one, as expected. After the convolution of HARPS spectrum to the respective resolution, the overall trend of their values become highly compatible.

\begin{table*}
\centering
\caption{The pseudo EWs of three absorption lines for HARPS spectrum Gl176 in different resolutions. The lower the resolution, the smaller the pseudo EW.}
\begin{tabular}{cccc}
\hline\hline\\
Resolution & p.EW of $\lambda$ 6536.67 & p.EW of $\lambda$ 6537.08 & p.EW of $\lambda$ 6537.64 \\
 & [m$\AA$] & [m$\AA$]  & [m$\AA$] \\
\hline\\
original 115000 & 14.19  & 29.08  & 27.31  \\
convolved 110000 & 14.04  & 28.89  & 26.98  \\
convolved 94600 & 13.39  & 28.27  & 26.35  \\
convolved 75000 & 11.99  & 27.45  & 25.08  \\
convolved 48000 & 7.50  & 22.50  & 20.04 \\
\hline\\
\end{tabular}
\label{EWvalues}
\end{table*}

\subsection{Measurements on different spectrographs}
\label{s2}

We examine the performance of our tool in new spectra. 
We show the accuracy of the stellar parameters predicted and the precision for each resolution, by presenting the mean absolute errors of the models and the dispersion of the results, as calculated after the 100 determinations for each spectrum.

For the case of HARPS, we use a HARPS spectrum of Gl643 with SNR = 83, which is not part of the HARPS dataset used in the machine learning. As reference values for this star, we consider its parameters reported by \citet{neves14}.
For the cases of the other instruments, we use a UVES spectrum of Gl846 with SNR = 149, a CARMENES spectrum of Gl514 with SNR = 191, a SOPHIE spectrum of Gl908 with SNR = 90 and a FEROS spectrum of Gl674 with SNR = 61. As reference values to those spectra, we consider the values of the respective HARPS ones in the dataset.

The results of T$_{\mathrm{eff}}$ and [Fe/H] are presented in Table~\ref{newstarspar}. We notice that the parameters of the new spectra are very close to the respective reference values. The differences in T$_{\mathrm{eff}}$ vary up to $\sim$50 K and the differences in [Fe/H] vary up to 0.03 dex. The mean absolute errors of models and the dispersions of values are slightly growing towards lower resolutions.

\subsection{Measurements on different SNR's}
\label{snr}

Here we examine the possible variation of the results regarding different signal-to-noise ratios (SNR) for a given spectrum.
We take the spectrum Gl514 of CARMENES, which has the highest SNR of the ones we examine (equal to 191, as reported in the CARMENES data archive) and we inject amounts of noise which correspond to lower SNR values that we set. Since the final noise is obtained by the quadratic sum of the initial noise and the injected noise, the final SNR values are calculated using the relation below.
\begin{equation}
(\frac{1}{SNR})^{2}_{final}=(\frac{1}{SNR})^{2}_{initial} + (\frac{1}{SNR})^{2}_{injected}
\end{equation}

\begin{figure}
\centering
$\begin{array}{c}
\includegraphics[width=9cm]{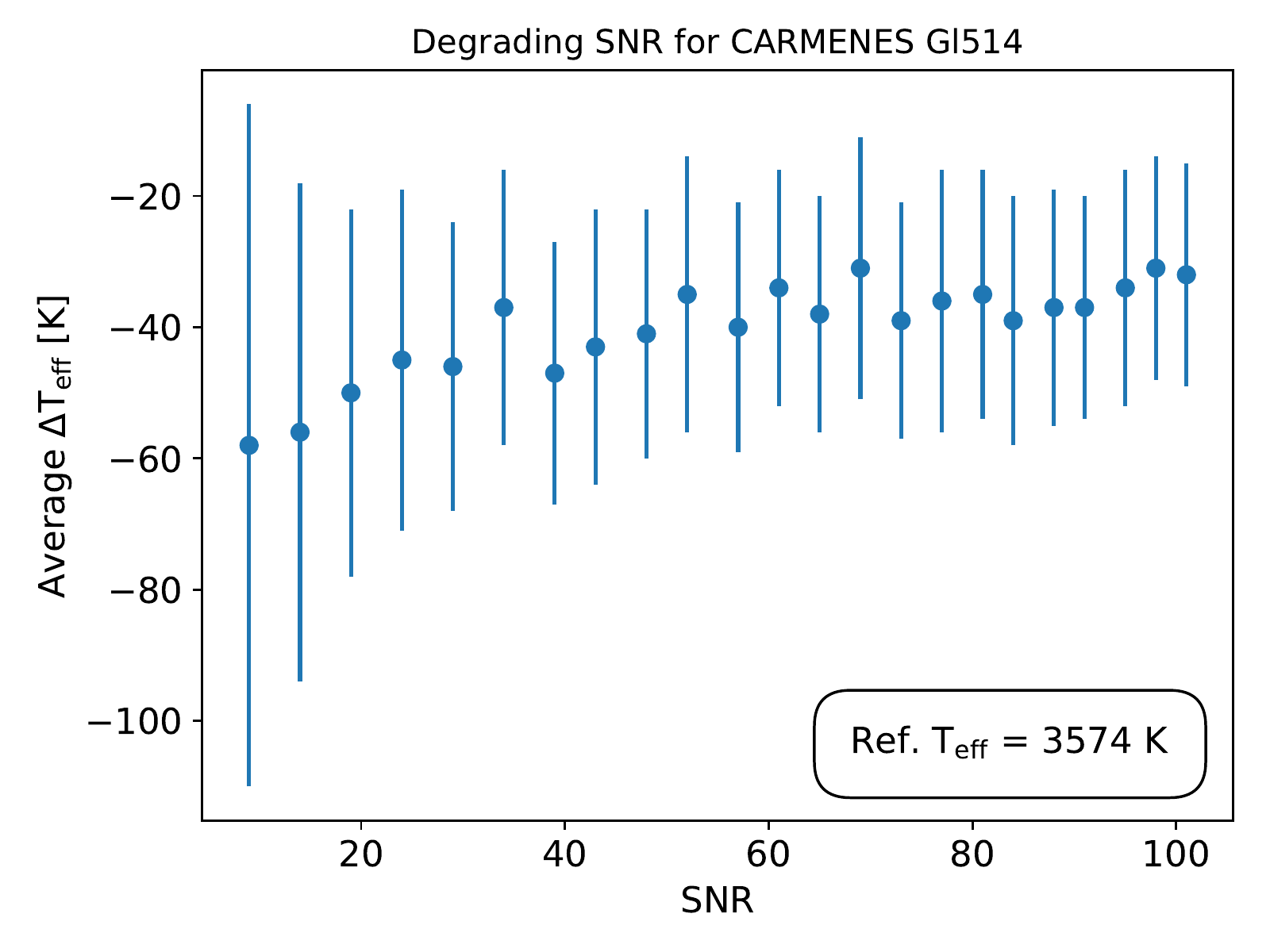} \\
\includegraphics[width=9cm]{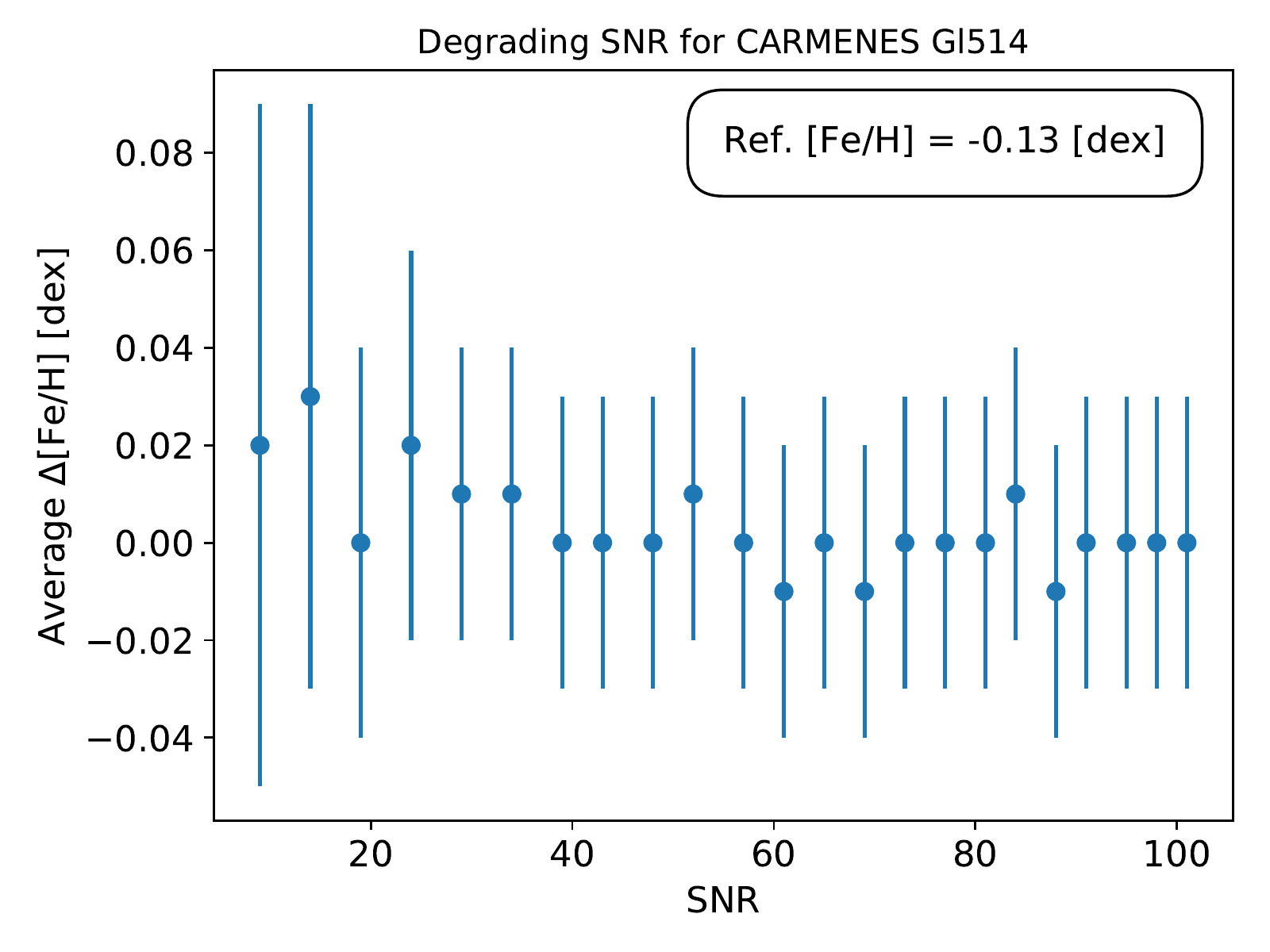} \\
\end{array}$
\caption{The average differences and the dispersion of T$_{\mathrm{eff}}$ (upper panel) and [Fe/H] (lower panel) for CARMENES Gl514 spectrum with different values of SNR.}
\label{snrcomp} 
\end{figure}

We create new spectra with final SNR values ranging from 100 to 9. 
For each spectrum, we measure the stellar parameters and their dispersion.
Fig.~\ref{snrcomp} illustrates the measurements of the CARMENES spectrum while degrading its SNR.
Overall, the results are similar to the ones of the original spectrum and the differences are kept roughly constant with respect to the reference values. 
For SNR values down to 20, we notice that the dispersions are between 17 and 27 K for T$_{\mathrm{eff}}$ and between 0.03 and 0.04 dex for [Fe/H], i.e. at similar levels as those of the original spectrum. 
For SNR values below 20, the dispersions start to increase up to  $\sim$50 K and up to $\sim$0.07 dex respectively. 
Moreover, it seems that there is a slight decrease of the order of 20 K in T$_{\mathrm{eff}}$ and a slight increase of the order of 0.02 dex in [Fe/H] for the spectra with SNR below 20. However,  these results are within the uncertainties of the tool.
Therefore, we conclude that our tool works consistently for spectra with SNR above 20. Bellow this SNR, the errors increase significantly.

\subsection{Comparison of results between our tool and \citet{neves14}}
\label{oduvn}

Now, we make an overall comparison of our results on a group of HARPS spectra with the ones presented by \citet{neves14}.
For this purpose, we measure 30 HARPS spectra from the initial GTO sample,  for which we do not know their parameters from photometry and are not part of the machine learning dataset we use. Based on the information from \citet{neves14}, we have excluded very active stars and stars with SNR lower than 25, below which that method does not apply. 
Both methods have been tested and do not work properly for very active or young stars, since the pseudo EWs of such spectra are affected and their parameters can not be determined accurately with the pseudo-EW approach we follow.
Then, we compare the results we get by our tool with the results presented by \citet{neves14}. 

The errors of the stellar parameters derived using our tool, are 27 K for T$_{\mathrm{eff}}$ and 0.04 dex for [Fe/H], as the mean absolute errors are measured when the machine learning model is applied on the test dataset. The errors of the calibration by \citet{neves14}, which are quantified from the root mean squared error (RMSE) in that work, are equal to 91 K and 0.08 dex respectively. It is reminded that both methods are tied to the same initial systematic uncertainties of the reference parameters used, which are 100K for T$_{\mathrm{eff}}$ and 0.17 dex for [Fe/H]. 

\begin{figure}
\centering
$\begin{array}{c}
\includegraphics[width=\hsize]{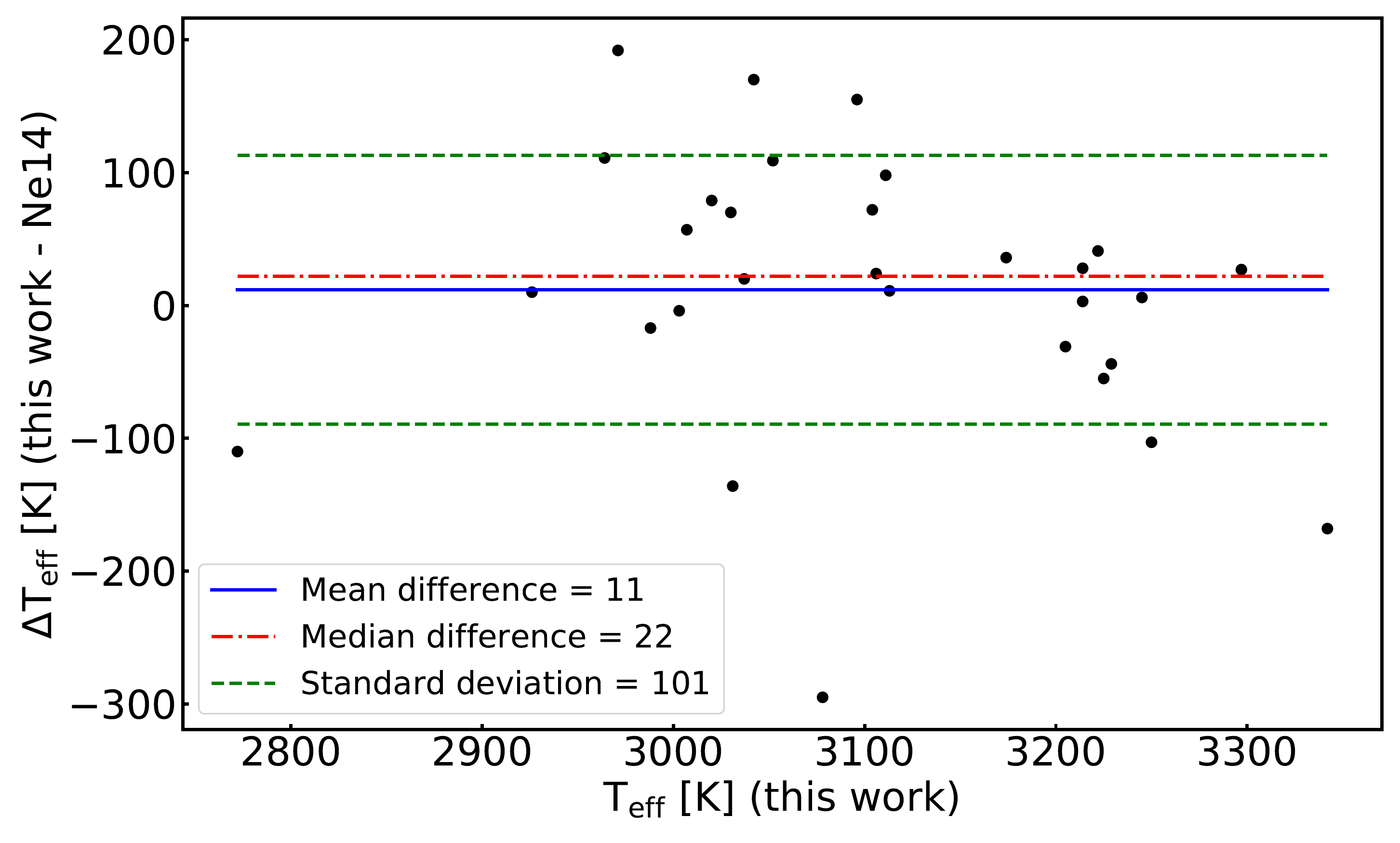} \\
\includegraphics[width=\hsize]{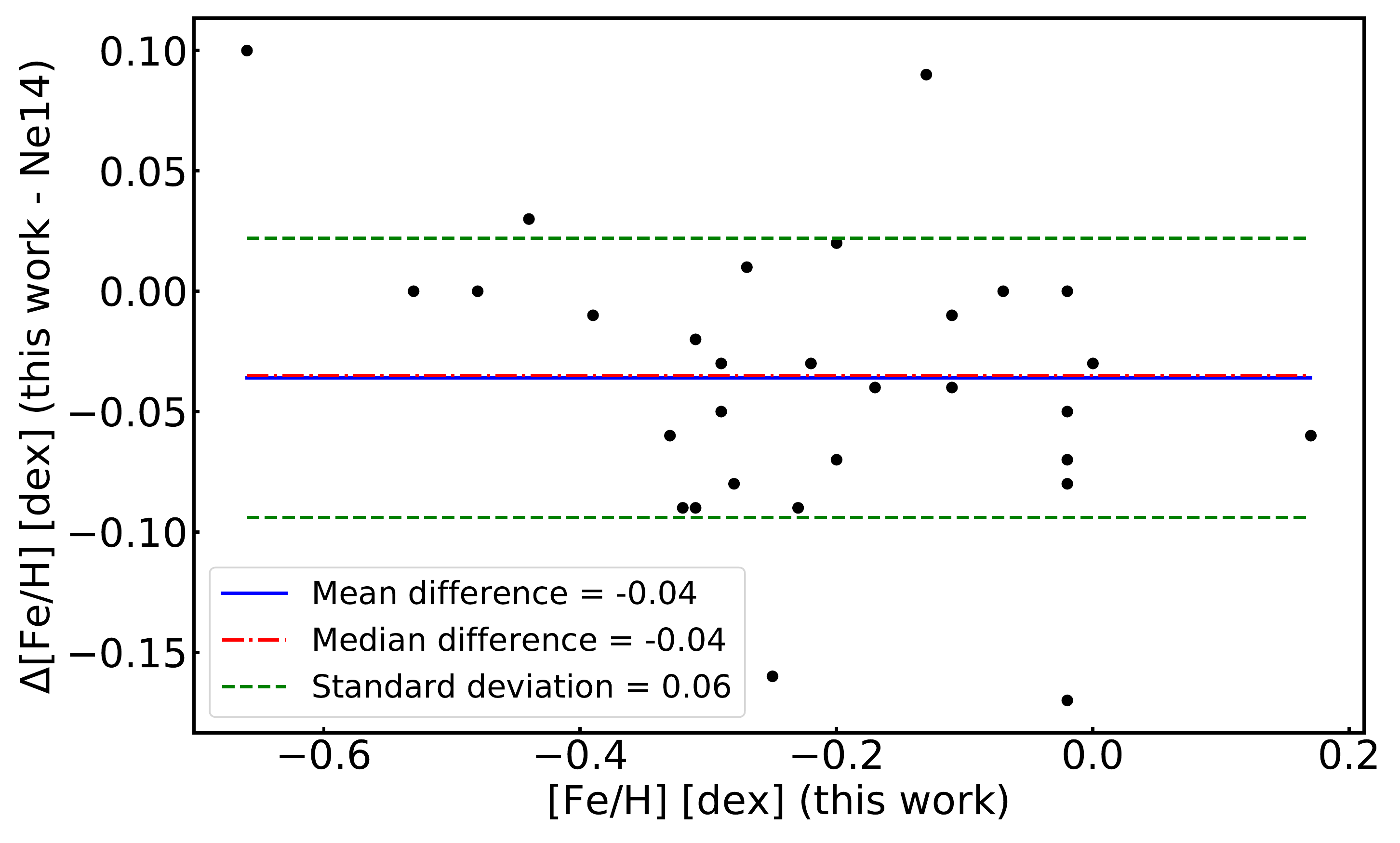}\\
\end{array}$
\caption{T$_{\mathrm{eff}}$ comparison (upper panel) and [Fe/H] comparison (lower panel) between this work and \citet{neves14}.}
\label{paracomp} 
\end{figure}

The results and their differences are presented in Table~\ref{parall} and Fig.~\ref{paracomp}. 
The mean and median difference of T$_{\mathrm{eff}}$ is 11 and 22 K respectively, with a standard deviation of 101 K. Regarding [Fe/H], the mean and median difference is -0.04 dex, with a standard deviation of 0.06 dex. 

Work by \citet{neves14} follows a traditional approach, using a least-squares weighted fit to determine parameters. The regression of our tool reduces those errors of T$_{\mathrm{eff}}$ and [Fe/H] from 91 to 27 K and from 0.08 to 0.04 dex respectively. So, our machine learning approach increases significantly the precision of parameter determinations. 
In terms of speed, the determination for a star by machine learning, even after the multiple runs with shuffling and splitting again the train/test samples, is a matter of few seconds.

\begin{table*}
\centering
\caption{Stellar parameters of 30 HARPS spectra as calculated by our tool (AA), by \citet{neves14} (Ne14) and their difference. The SNR of those stars are between 28 and 97, as reported by \citet{neves14}.}
\begin{tabular}{ccccccc}
\hline\hline\\
Star &  T$_{\mathrm{eff}}$ (AA) & T$_{\mathrm{eff}}$ (Ne14) & T$_{\mathrm{eff}}$ Diff. & [Fe/H] (AA) & [Fe/H] (Ne14) & [Fe/H] Diff.  \\ 
  &  [$\pm$27 K] & [$\pm$91 K] & [K] & [$\pm$0.04 dex] & [$\pm$0.08 dex] & [dex] \\ 
\hline\\
CD-44-836A & 3104 & 3032 & 72 & -0.07 & -0.07 & 0.00 \\
G108-21 & 3214 & 3186 & 28 & -0.02 & -0.02 & 0.00 \\
GJ1057 & 2926 & 2916 & 10 & -0.11 & -0.10 & -0.01 \\
GJ1061 & 2772 & 2882 & -110 & -0.25 & -0.09 & -0.16 \\
GJ1065 & 3106 & 3082 & 24 & -0.32 & -0.23 & -0.09 \\
GJ1123 & 2971 & 2779 & 192 & -0.02 & 0.15 & -0.17 \\
GJ1129 & 3037 & 3017 & 20 & -0.02 & 0.05 & -0.07 \\
GJ1236 & 3225 & 3280 & -55 & -0.44 & -0.47 & 0.03 \\
GJ1256 & 2964 & 2853 & 111 & -0.02 & 0.06 & -0.08 \\
GJ1265 & 3020 & 2941 & 79 & -0.28 & -0.20 & -0.08 \\
Gl12 & 3245 & 3239 & 6 & -0.31 & -0.29 & -0.02 \\
Gl145 & 3297 & 3270 & 27 & -0.27 & -0.28 & 0.01 \\
Gl203 & 3174 & 3138 & 36 & -0.31 & -0.22 & -0.09 \\
Gl299 & 3078 & 3373 & -295 & -0.53 & -0.53 & 0.00 \\
Gl402 & 3052 & 2943 & 109 & 0.00 & 0.03 & -0.03 \\
Gl480.1 & 3214 & 3211 & 3 & -0.48 & -0.48 & 0.00 \\
Gl486 & 3096 & 2941 & 155 & -0.02 & 0.03 & -0.05 \\
Gl643 & 3113 & 3102 & 11 & -0.29 & -0.26 & -0.03 \\
Gl754 & 2988 & 3005 & -17 & -0.23 & -0.14 & -0.09 \\
L707-74 & 3250 & 3353 & -103 & -0.39 & -0.38 & -0.01 \\
LHS1134 & 3007 & 2950 & 57 & -0.20 & -0.13 & -0.07 \\
LHS1481 & 3342 & 3510 & -168 & -0.66 & -0.76 & 0.10 \\
LHS1723 & 3031 & 3167 & -136 & -0.29 & -0.24 & -0.05 \\
LHS1731 & 3229 & 3273 & -44 & -0.22 & -0.19 & 0.03 \\
LHS1935 & 3222 & 3181 & 41 & -0.20 & -0.22 & 0.02 \\
LHS337 & 3003 & 3007 & -4 & -0.33 & -0.27 & -0.06 \\
LHS3583 & 3205 & 3236 & -31 & -0.13 & -0.22 & 0.09 \\
LHS3746 & 3111 & 3013 & 98 & -0.17 & -0.13 & -0.04 \\
LHS543 & 3042 & 2872 & 170 & 0.17 & 0.23 & -0.06 \\
LP816-60 & 3030 & 2960 & 70 & -0.11 & -0.07 & -0.04 \\
\hline\\
\end{tabular}
\label{parall}
\end{table*}

\subsection{Estimating total uncertainties }
\label{gauss}

Intrinsic uncertainties exist in the T$_{\mathrm{eff}}$ and [Fe/H] reference values of the HARPS dataset, since their initial photometric derivations have average uncertainties of 100 K and 0.17 dex respectively. 
Since these parameters are used as the training values for the machine learning process, we decide to inject these uncertainties by perturbing their values accordingly, in order to see how the final results of the predictions will vary.

Therefore, we create gaussian distributions on the parameters for each HARPS training dataset, increasing the dispersion of distribution on the reference parameters each time with step of 10 K and 0.02 dex, until the uncertainties of 100 K and 0.17 dex. This adds different training values to the machine learning algorithm each time.
For each step, we create 100 gaussian-distributed training datasets. After these runs of machine learning, we calculate the average values of predicted parameters and their dispersion. 

\begin{figure*}
\centering
$\begin{array}{cc}
\includegraphics[width=9cm]{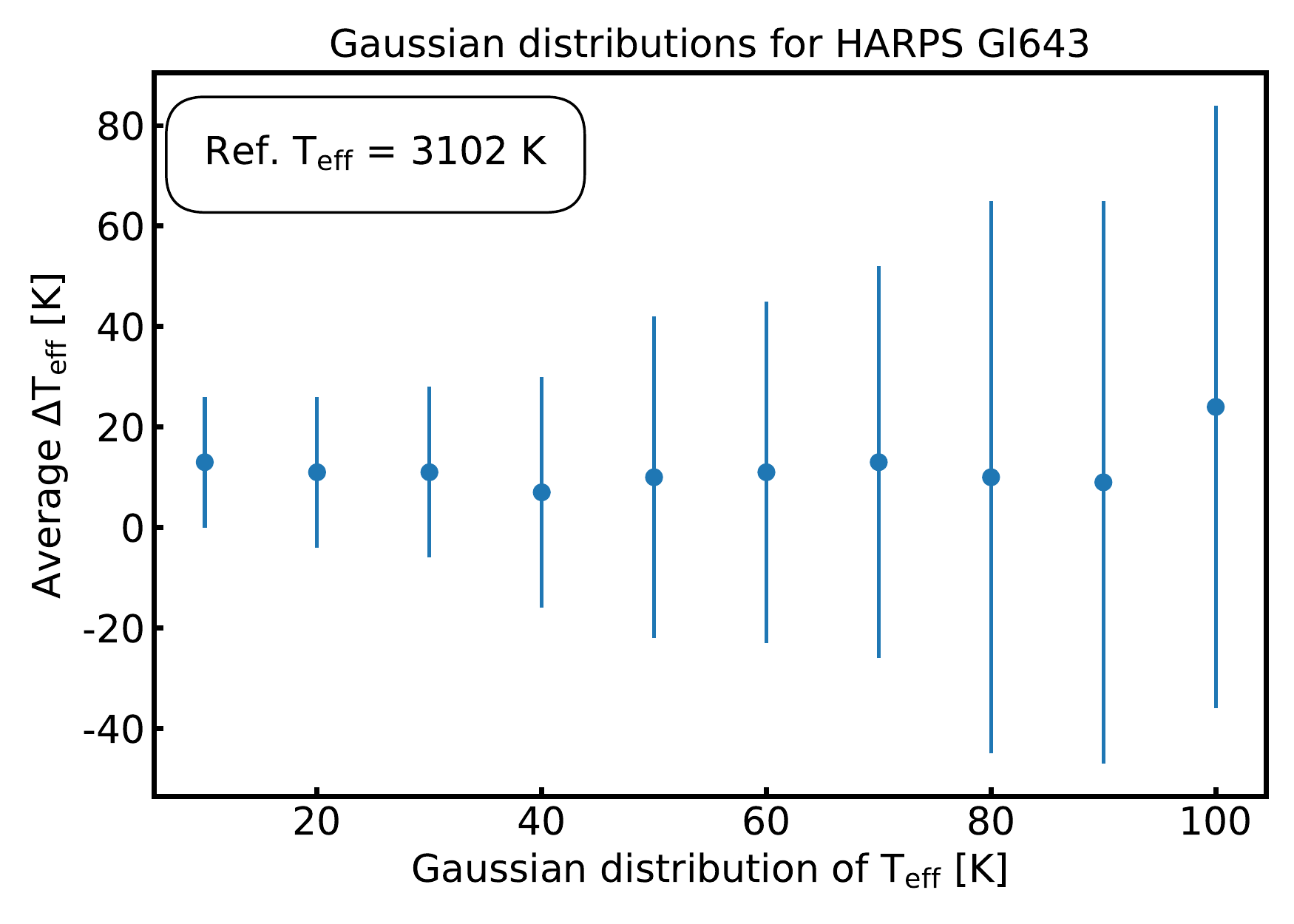} &
\includegraphics[width=9cm]{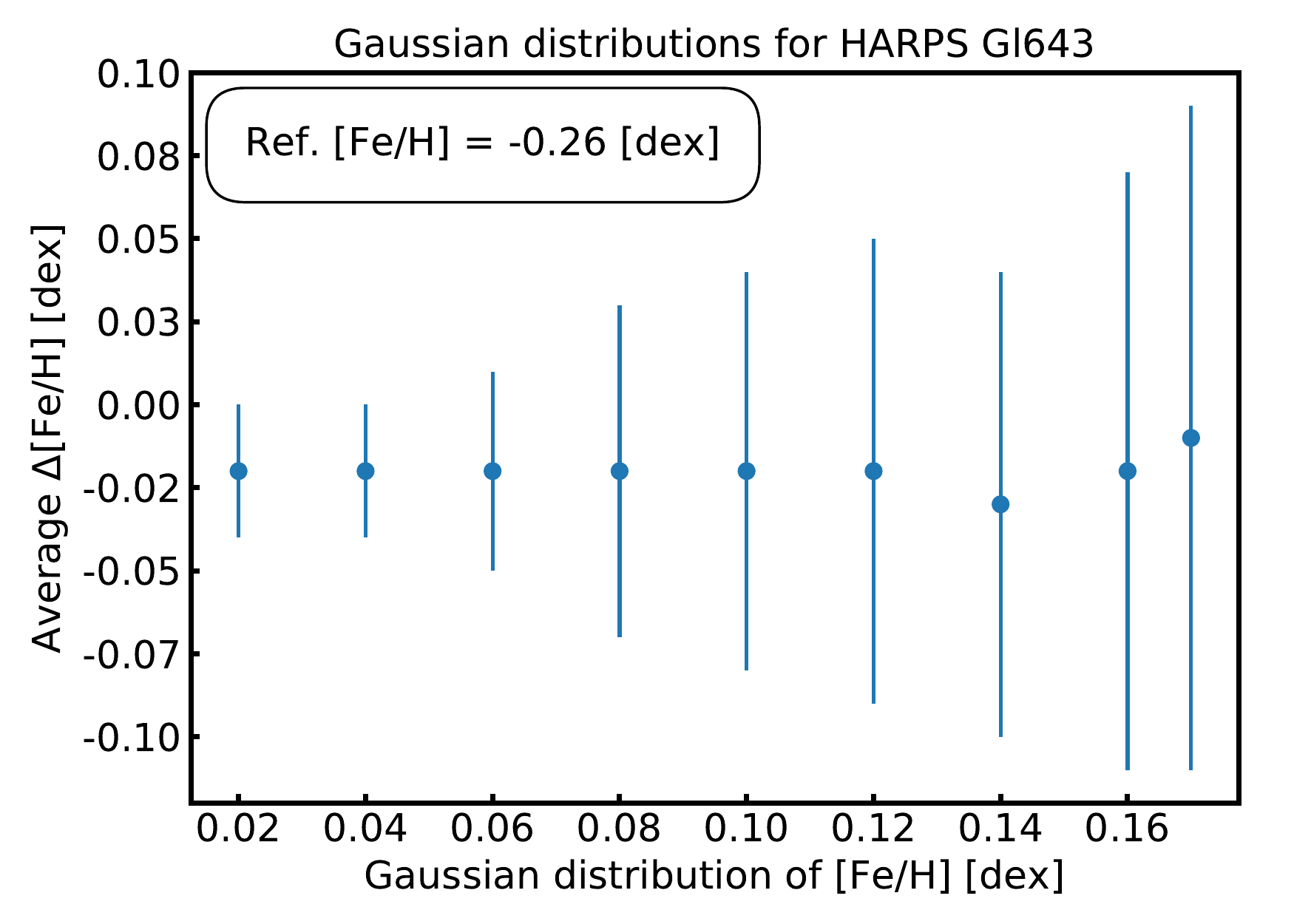} \\
\includegraphics[width=9cm]{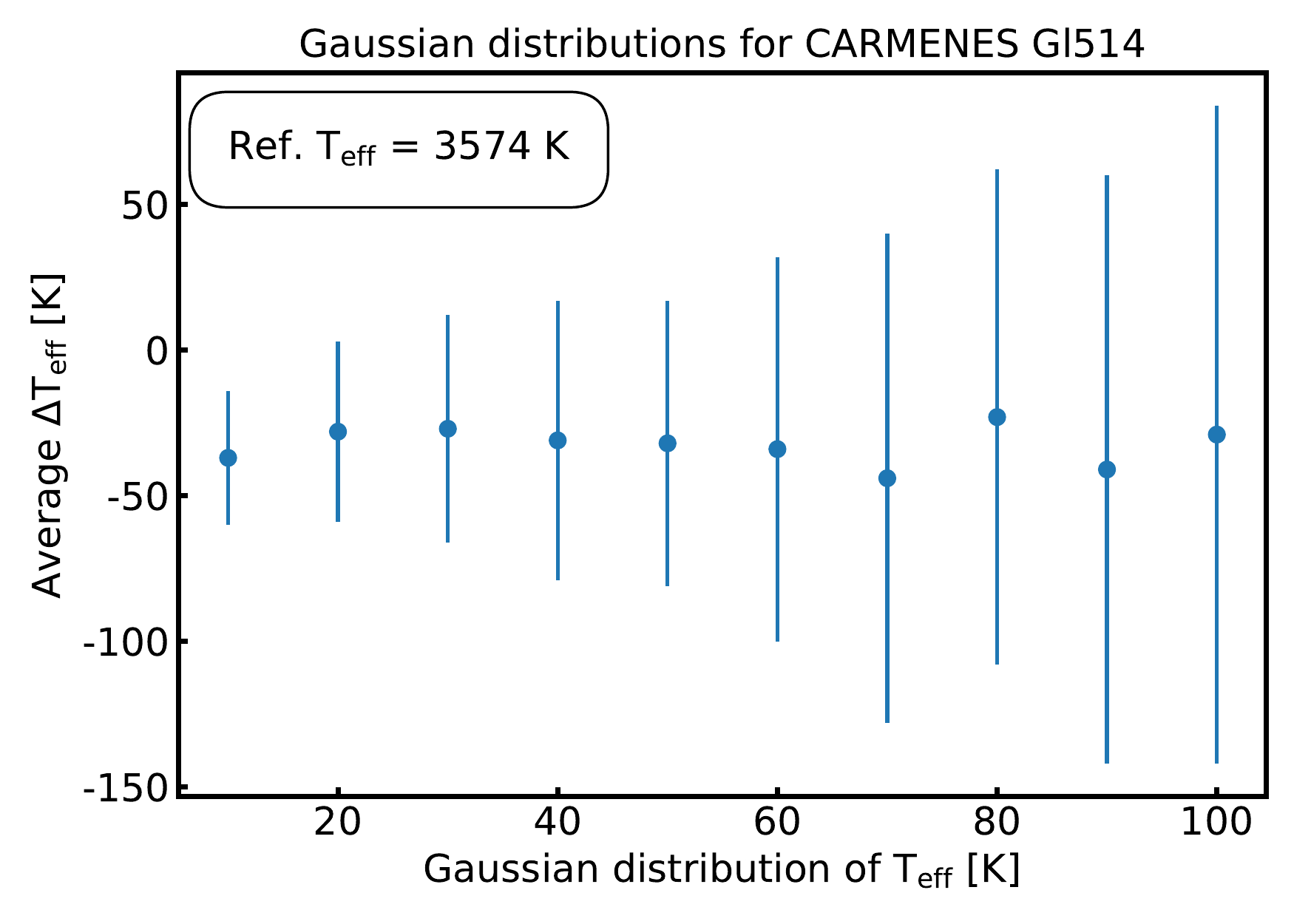} & 
\includegraphics[width=9cm]{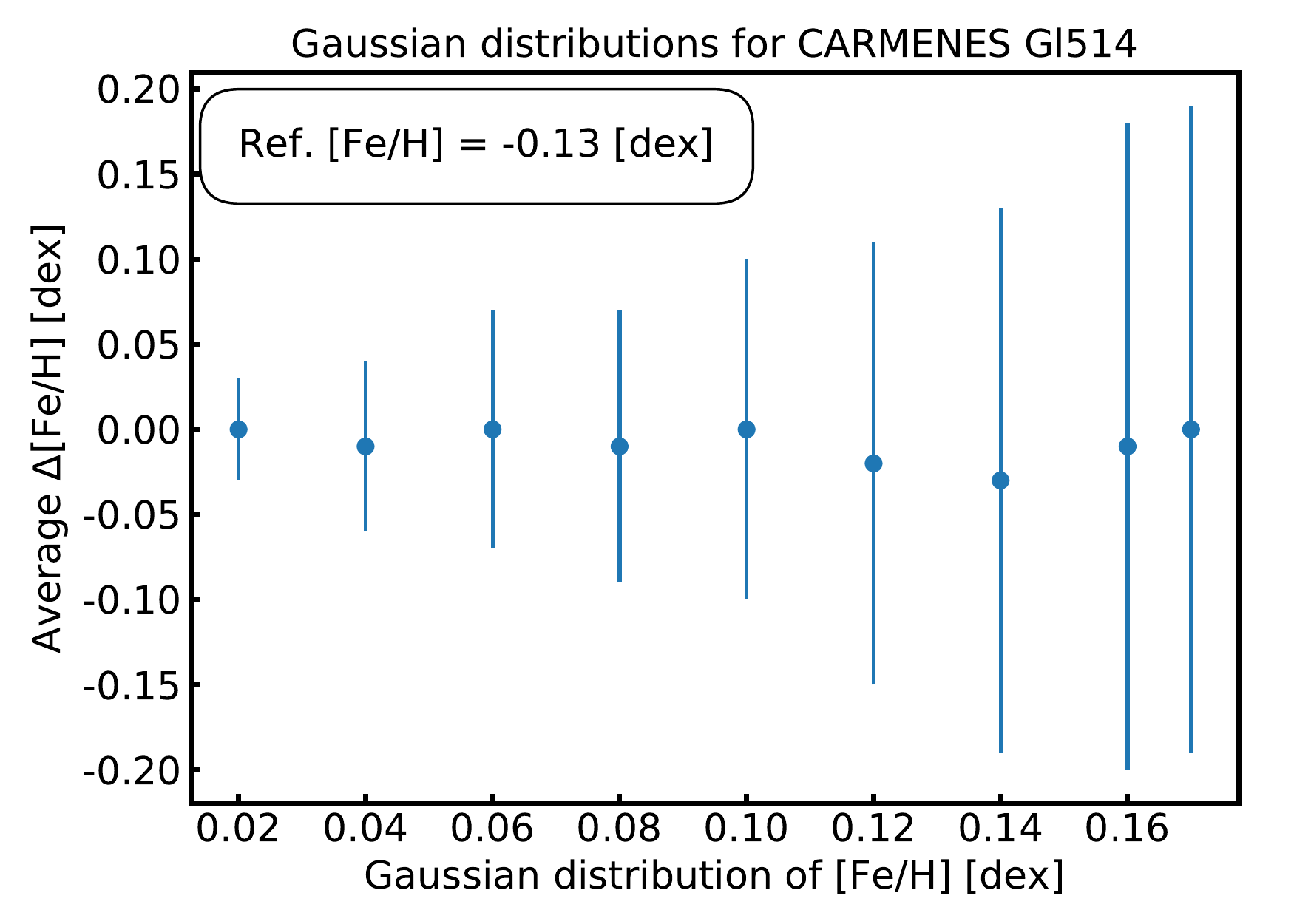}  \\
\includegraphics[width=9cm]{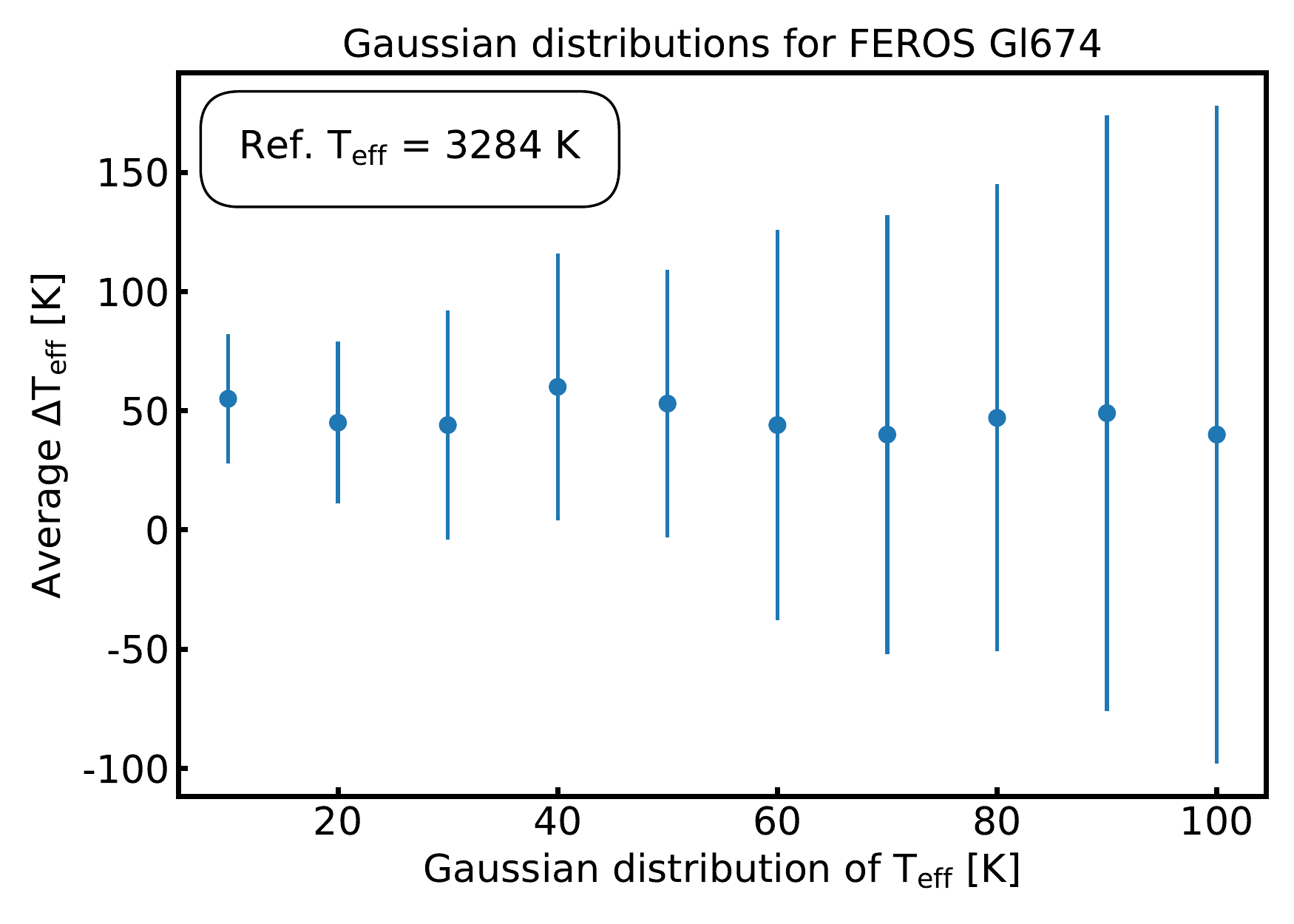} & 
\includegraphics[width=9cm]{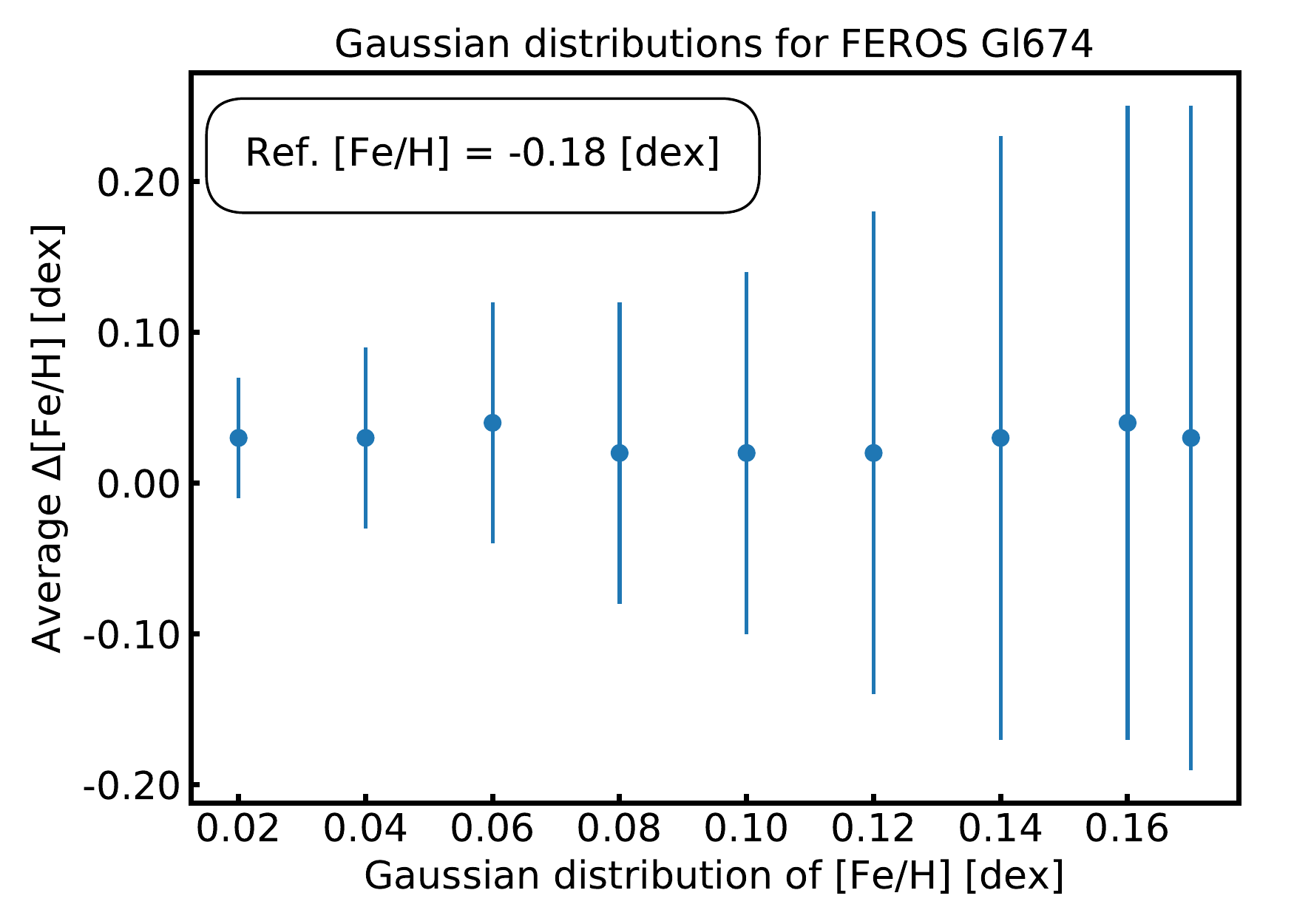}  \\

\end{array}$
\caption{The average differences and the dispersions of the results for several amounts of gaussian distribution injected to the reference parameters of the training HARPS datasets. The result of each step is the average outcome from 100 different distributed datasets.}
\label{gd} 
\end{figure*}

In Fig.~\ref{gd}, we present the variations for spectra from the highest resolution (HARPS), the lowest resolution (FEROS) and an intermediate resolution (CARMENES). 
The datapoints represent the average difference between the resulting parameters and the reference values, after being calculated with the 100 different datasets. The errorbars are the dispersion of it.
We notice that the average differences from the reference values are almost the same among them, regardless the amount of uncertainty injected to the gaussian distribution. 

The average results of T$_{\mathrm{eff}}$ and [Fe/H] for the spectra from all the instruments are presented in Table~\ref{gdstarpar}. We report their maximum errors after considering the maximum gaussian distribution with 100 K and 0.17 dex. Overall, the average values of the parameters remain roughly the same as the ones calculated with no gaussian distribution at all. 
The mean absolute errors (M.A.E.) of the machine learning models have grown to values between 65 and 80 K for T$_{\mathrm{eff}}$ and between 0.10 to 0.13 dex for [Fe/H], depending on the resolution of the HARPS dataset.
The dispersion of the derived parameters grows as the resolution of the spectra becomes lower. 
Specifically, it is smaller than the injected uncertainties for the HARPS spectrum ($\sim$60 K and $\sim$0.10 dex), while for the spectra from other instruments, it is slightly higher than the uncertainties injected ($\sim$110 to $\sim$130 K and $\sim$0.18 to  $\sim$0.22 dex respectively).  

In all the cases though, the resulting average values of stellar parameters are very close to their expected values. Differences of T$_{\mathrm{eff}}$ are up to $\sim$40 K and differences of [Fe/H] are up to 0.03 dex, regarding to the expected values.

\begin{table*}
\centering
\caption{The machine learning (M.L.) results of T$_{\mathrm{eff}}$ and [Fe/H], their dispersion (Disp.), the mean absolute errors (M.A.E.) of the models and the reference values (Ref.) for comparison.}
\begin{tabular}{ccccccccccc}
\hline\hline\\
Star & Spec. & Res. & Ref. & M.L.  & M.A.E.  & Disp. & Ref. & M.L. & M.A.E. & Disp. \\
 & & & T$_{\mathrm{eff}}$ & T$_{\mathrm{eff}}$ & T$_{\mathrm{eff}}$ & T$_{\mathrm{eff}}$ & [Fe/H] & [Fe/H] & [Fe/H] & [Fe/H]\\
 & & & [K] & [K] & [K] & [K] & [dex] & [dex] & [dex] & [dex] \\
\hline\\
Gl643 & HARPS & 115000 & 3102 & 3113 & 27 & 10 & -0.26  & -0.28 & 0.04 & 0.01 \\
Gl846  & UVES & 110000  &  3682 & 3691 & 28 & 13  & -0.08  & -0.05 & 0.04 & 0.02 \\
Gl514 & CARMENES & 94600 & 3574 & 3547 & 28 & 17 & -0.13 & -0.13 & 0.04 & 0.03 \\
Gl908 & SOPHIE & 75000 & 3587 & 3580 & 28 & 18 & -0.38 & -0.35 & 0.04 & 0.03 \\
Gl674 & FEROS & 48000 & 3284 & 3338 & 30 & 24 & -0.18  & -0.16 & 0.04 & 0.03 \\
\hline\\
\end{tabular}
\label{newstarspar}
\end{table*}

\begin{table*}
\centering
\caption{The machine learning (M.L.) results of T$_{\mathrm{eff}}$ and [Fe/H] after injecting uncertainties with gaussian distributions of 100 K and 0.17 dex in the parameters of the training HARPS datasets, their dispersion (Disp.), the mean absolute errors (M.A.E.) of the models and the reference values (Ref.) for comparison.}
\begin{tabular}{ccccccccccc}
\hline\hline\\
Star & Spec. & Res. & Ref. & M.L.  & M.A.E.  & Disp. & Ref. & M.L. & M.A.E. & Disp. \\
 & & & T$_{\mathrm{eff}}$ & T$_{\mathrm{eff}}$ & T$_{\mathrm{eff}}$ & T$_{\mathrm{eff}}$ & [Fe/H] & [Fe/H] & [Fe/H] & [Fe/H]\\
 & & & [K] & [K] & [K] & [K] & [dex] & [dex] & [dex] & [dex] \\
\hline\\
Gl643 & HARPS & 115000 & 3102 & 3126 & 65 & 60 & -0.26  & -0.29 & 0.10 & 0.10 \\
Gl846  & UVES & 110000  &  3682 & 3678 & 68 & 109 & -0.08  & -0.06 & 0.10 & 0.18 \\
Gl514 & CARMENES & 94600 & 3574 & 3545 & 77 & 113 & -0.13 & -0.13 & 0.12 & 0.19 \\
Gl908 & SOPHIE & 75000 & 3587 & 3585 & 78 & 120 & -0.38 & -0.36 & 0.13 & 0.21 \\
Gl674 & FEROS & 48000 & 3284 & 3324 & 80 & 138 & -0.18  & -0.15 & 0.13 & 0.22 \\
\hline\\
\end{tabular}
\label{gdstarpar}
\end{table*}

\subsection{Validation of [Fe/H] determinations by measuring binary systems}
\label{binaries}

Here, we measure [Fe/H] in binary systems, containing M dwarfs which are not part of the reference sample used for machine learning. Thus, we validate our method of [Fe/H] prediction in an independent way. We determine [Fe/H] both in FGK+M and in M+M systems for an even more intrinsic test of [Fe/H] agreement.

The [Fe/H] determinations of eight FGK+M binary systems, from spectra obtained by UVES and FEROS spectrographs, are presented in Table~\ref{FGKM}. 
Regarding the FGK stars, their [Fe/H] and respective uncertainties were derived using the methodology described in \citet{sousa} and \citet{santos13}. 
The method measures the equivalent widths of FeI and FeII lines and assumes ionization and excitation equilibrium. 
It makes use of the radiative transfer code MOOG \citep{sneden} and a grid of Kurucz model atmospheres \citep{kurucz}.
The [Fe/H] values  of the respective M dwarf secondaries, derived by ODUSSEAS, are presented along with the total uncertainties of our tool at the resolutions of UVES (0.10 dex) and FEROS (0.13 dex).
All binaries have differences within the uncertainties of the methods.

Furthermore, we proceed to [Fe/H] determinations of stars in five M+M binary systems, measuring their available spectra from the CARMENES public archive.
In Table~\ref{MM}, we present these results along with their own dispersions, since both are estimated by our tool based on the same reference values with the same initial uncertainties.
We notice agreement between the respective members of all the M+M binaries, within the dispersions of their [Fe/H] determinations.
This is a validation that our tool predicts [Fe/H] in a consistent and accurate way.

\begin{table*}
\centering
\caption{[Fe/H] difference between members of FGK+M binary systems.}
\begin{tabular}{ccccccc}
\hline\hline\\
Primary & [Fe/H] &  $\sigma$[Fe/H] & Secondary &  [Fe/H] &  $\sigma$[Fe/H] & [Fe/H] Difference \\
 & [dex] & [dex]  & & [dex] & [dex] & [dex] \\
\hline\\
Gl100A & -0.29 & 0.05  & Gl100C & -0.29 & 0.13 & 0.00 \\
Gl118.1A & 0.02  & 0.05  & Gl118.1B & 0.05 & 0.10 & 0.03 \\
Gl173.1A & -0.37 & 0.03 & Gl173.1B  & -0.29 & 0.10 & 0.08 \\
Gl157A & -0.08  & 0.04  & Gl157B  & 0.02 & 0.13 & 0.10 \\
NLTT19073 & 0.08  & 0.03  & NLTT19072 & -0.07 & 0.13 & -0.15 \\
NLTT29534 & 0.00  & 0.03  & NLTT29540 & -0.03 & 0.10 & -0.03 \\
NLTT34137 & -0.12  & 0.05  & NLTT34150 & -0.13 & 0.10 & -0.01\\
NLTT34353 & -0.10  & 0.03  & NLTT34357 & -0.11 & 0.10 & -0.01\\
\hline\\
\end{tabular}
\label{FGKM}
\end{table*}

\begin{table*}
\centering
\caption{[Fe/H] difference between members of M+M binary systems.}
\begin{tabular}{ccccccc}
\hline\hline\\
Primary & [Fe/H] &  Disp. & Secondary &  [Fe/H] &  Disp. & [Fe/H] Difference \\
 & [dex] & [dex]  & & [dex] & [dex] & [dex] \\
\hline\\
Gl553 & -0.07 & 0.06  & Gl553.1 & -0.10 & 0.07 & 0.03 \\
Gl875 & -0.15  & 0.05  & Gl875.1 & -0.17 & 0.06 & 0.02 \\
Gl617A & 0.04  & 0.04  & Gl617B & -0.08 & 0.08 & 0.12 \\
Gl745A & -0.48 & 0.04 & Gl745B  & -0.53 & 0.06 & 0.05 \\
Gl752A & 0.01  & 0.03  & Gl752B  & -0.07 & 0.08 & 0.08 \\
\hline\\
\end{tabular}
\label{MM}
\end{table*}

\subsection{Discussion on the reference parameter scales}
\label{referencescales}

\begin{figure}
\centering
$\begin{array}{c}
\includegraphics[width=\hsize]{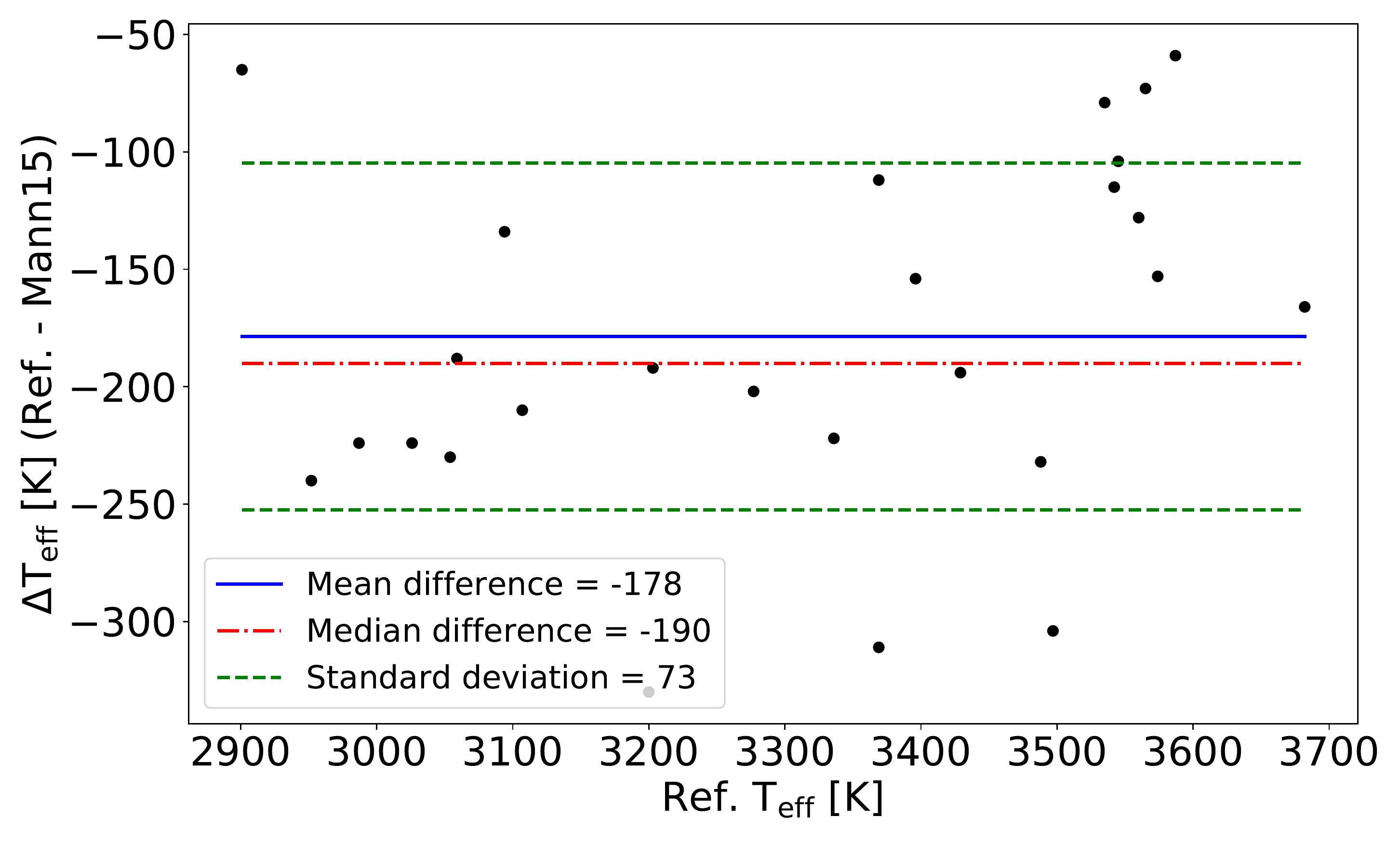} \\
\includegraphics[width=\hsize]{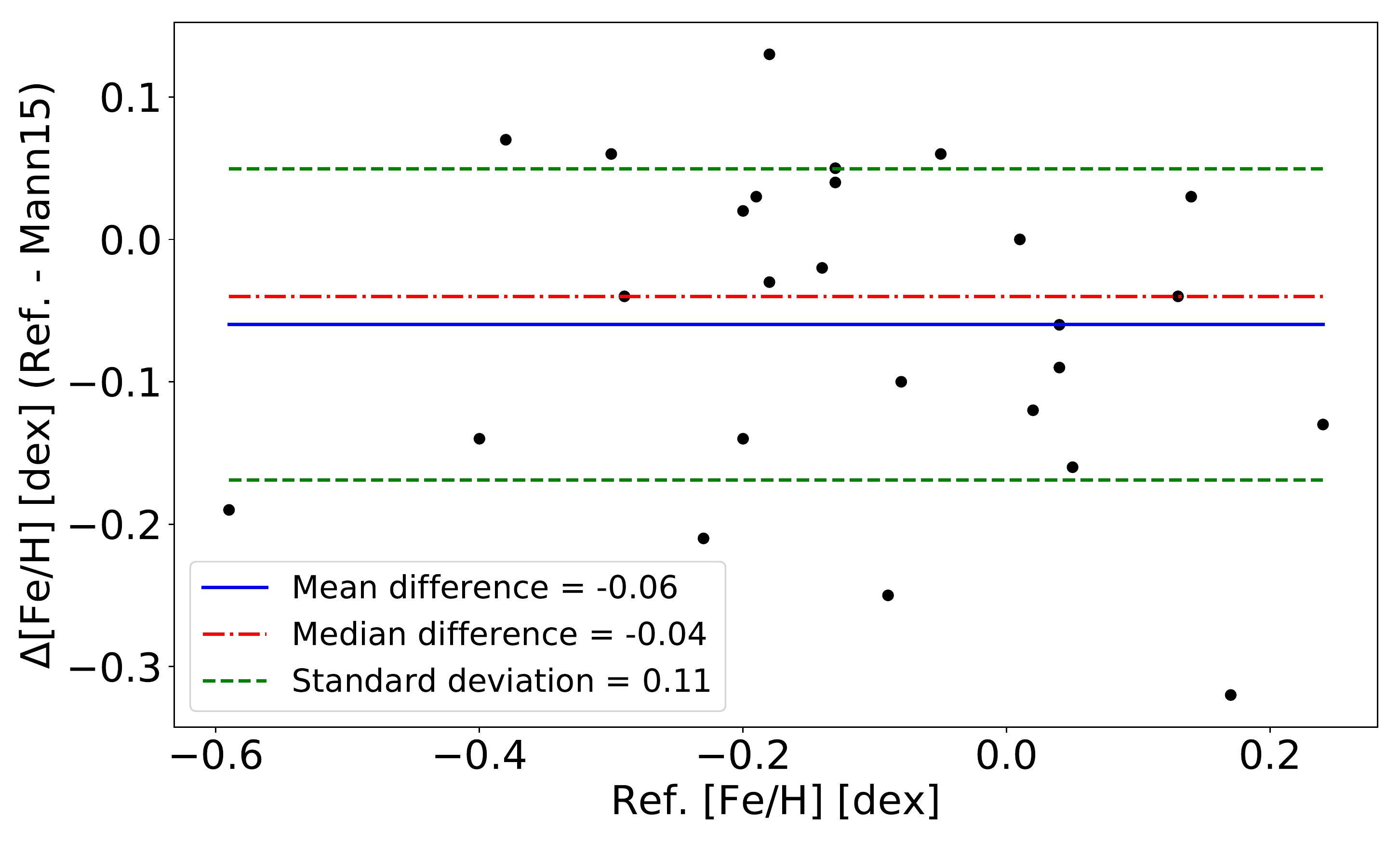}\\
\end{array}$
\caption{T$_{\mathrm{eff}}$ comparison (upper panel) and [Fe/H] comparison (lower panel) between the reference values we use and \citet{mann15} for 26 common stars.}
\label{scalecomp} 
\end{figure}

Since supervised machine learning determines the parameters based on reference values given to it, their systematics will apply to the results of new stars too.
In this work, we have used the reference  T$_{\mathrm{eff}}$ and [Fe/H]  photometric scales of \citet{casag08} and \citet{neves12} respectively, as they are derived in a homogeneous way for a sufficiently big number of spectra available to us.
It is important to make a comparison between the reference values we use and values of same stars derived by other recent works, which may be subject to different systematics. Such is \citet{mann15}, with which we share 26 common stars of the 65 ones we use as our reference dataset.
In Table~\ref{scales}, we compare our reference parameters with determinations by \citet{mann15} and report the differences.
These differences are illustrated in Figure~\ref{scalecomp}.
Regarding T$_{\mathrm{eff}}$, we notice that our reference values have a systematic underestimation of 178 K on average with a standard deviation of 73 K. This systematic difference roots back to the different methods of derivation followed. Work by  \citet{casag08} is based on the multiple optical-infrared technique (MOITE) for M dwarfs, which is an extension of the infrared flux method (IRFM) as described in \citet{casag06}. On the other hand, determinations by \citet{mann15} are done by comparing the optical spectra with the CFIST suite of the BT-SETTL version of the PHOENIX atmosphere models \citep{allard13}. The detailed description of this method can be found in  \citet{mann13b}.
Regarding [Fe/H], we notice no significant systematic difference between the methods of calibration by \citet{neves12} and \citet{mann15}. The average difference is 0.06 dex with a standard deviation of 0.11 dex for the sample of stars in common.

As a potential future improvement of our determinations, we consider the possibility of replacing our reference dataset. 
Since new techniques of parameter determination become more accurate and precise and as more spectra will become available to us, their homogeneously derived parameters can be correlated with their pseudo EWs. 
Thus, we take into account the creation of an improved reference dataset for our machine learning tool.

\begin{table*}
\centering
\caption{Stellar parameters of 26 stars in common with \citet{mann15} and their difference.}
\begin{tabular}{ccccccc}
\hline\hline\\
Star &  T$_{\mathrm{eff}}$ (Ref.) & T$_{\mathrm{eff}}$ (Mann15) & T$_{\mathrm{eff}}$ Diff. & [Fe/H] (Ref.) & [Fe/H] (Mann15) & [Fe/H] Diff.  \\ 
  &  [$\pm$100 K] & [$\pm$60 K] & [K] & [$\pm$0.17 dex] & [$\pm$0.08 dex] & [dex] \\ 
\hline\\
Gl54.1 & 2091 & 3056 & -65 & -0.40 & -0.26 & -0.14 \\
Gl87 & 3565 & 3638 & -73 & -0.30 & -0.36 & 0.06 \\
Gl105B & 3054 & 3284 & -230 & -0.14 & -0.12 & -0.02 \\
Gl176 & 3369 & 3680 & -311 & 0.02 & 0.14 & -0.12 \\
Gl205 & 3497 & 3801 & -304 & 0.17 & 0.49 & -0.32 \\
G213 & 3026 & 3250 & -224 & -0.19 & -0.22 & -0.03 \\
Gl250B & 3369 & 3481 & -112 & -0.09 & 0.14 & -0.25 \\
Gl273 & 3107 & 3317 & -210 & -0.05 & -0.11 & 0.06 \\
Gl382 & 3429 & 3623 & -194 & 0.04 & 0.13 & -0.09 \\
Gl393 & 3396 & 3548 & -154 & -0.13 & -0.18 & 0.05 \\
Gl436 & 3277 & 3479 & -202 & 0.01 & 0.01 & 0.00 \\
Gl447 & 2952 & 3192 & -240 & -0.23 & -0.02 & -0.21 \\
Gl514 & 3574 & 3727 & -153 & -0.13 & -0.09 & -0.04 \\
Gl526 & 3545 & 3649 & -104 & -0.18 & -0.31 & 0.13 \\
Gl555 & 2987 & 3211 & -224 & 0.13 & 0.17 & -0.04 \\
Gl581 & 3203 & 3395 & -192 & -0.18 & -0.15 & -0.03 \\
Gl686 & 3542 & 3657 & -115 & -0.29 & -0.25 & -0.04 \\
Gl699 & 3094 & 3228 & -134 & -0.59 & -0.40 & -0.19 \\
Gl701 & 3535 & 3614 & -79 & -0.20 & -0.22 & 0.02 \\
Gl752A & 3336 & 3558 & -222 & 0.04 & 0.10 & -0.06 \\
Gl846 & 3682 & 3848 & -166 & -0.08 & 0.02 & -0.10 \\
Gl849 & 3200 & 3530 & -330 & 0.24 & 0.37 & -0.13 \\
Gl876 & 3059 & 3247 & -188 & 0.14 & 0.17 & -0.03 \\
Gl880 & 3488 & 3720 & -232 & 0.05 & 0.21 & -0.16 \\
Gl887 & 3560 & 3688 & -128 & -0.20 & -0.06 & -0.14 \\
Gl908 & 3587 & 3646 & -59 & -0.38 & -0.45 & -0.07 \\
\hline\\
\end{tabular}
\label{scales}
\end{table*}

\section{Summary}
\label{sum}

We present our machine learning tool ODUSSEAS for the derivation of T$_{\mathrm{eff}}$ and [Fe/H] in M dwarf stars, whose spectra can have different resolutions and wavelength ranges inside the area from 530 to 690 nm. We explain in detail the way it is built and works. We present the results of the tests we perform and we examine its accuracy and precision from very high resolution of 115000 to resolution of 48000. Our tool seems to be reliable, as it operates with high machine learning scores around 0.94 and achieves excellent predictions of significantly high precision with mean absolute errors of  $\sim$30 K for T$_{\mathrm{eff}}$ and $\sim$0.04 dex for [Fe/H]. Taking into consideration the intrinsic uncertainties of the reference parameters and perturbing them accordingly, our models have maximum uncertainties of $\sim$80 K for T$_{\mathrm{eff}}$ and $\sim$0.13 dex for [Fe/H], which are within the typical uncertainties for M dwarfs.
Our parameters for spectra from different spectrographs, occurring from the average of 100 determinations, have consistent values with differences within $\sim$50 K and $\sim$0.03 dex from the expected ones. Spectra should have SNR above 20 for optimal predictions.
Our tool is valid for M dwarfs in the intervals 2800 to 4000 K for T$_{\mathrm{eff}}$ and -0.83 to 0.26 dex for [Fe/H], except from very active or young stars.
 It can be tested by downloading the files in the webpage \url{https://github.com/AlexandrosAntoniadis/ODUSSEAS}, after reading the README instructions for clarifying the technical details.

\begin{acknowledgements}

This work was supported by FCT/MCTES through national funds and by FEDER - Fundo Europeu de Desenvolvimento Regional through COMPETE2020 - Programa Operacional Competitividade e Internacionalização by these grants: UID/FIS/04434/2019, UIDB/04434/2020 and UIDP/04434/2020; PTDC/FIS-AST/32113/2017 and POCI-01-0145-FEDER-032113; PTDC/FIS-AST/28953/2017 and POCI-
01-0145-FEDER-028953.
A.A.K., S.G.S., E.D.M. and G.D.C.T. acknowledge the support from FCT in the form of the
exploratory projects with references IF/00028/2014/CP1215/CT0002, IF/00849/2015/CP1273/CT0003 and IF/00956/2015/CP1273/CT0002.
S.G.S. and E.D.M. further acknowledge the support from FCT through the Investigador FCT
contracts IF/00028/2014/CP1215/CT0002, IF/00849/2015/CP1273/CT0003 and POCH/FSE (EC).
G.D.C.T. further acknowledges the support from an FCT/Portugal PhD grant with reference
PD/BD/113478/2015.

\end{acknowledgements}

\bibliographystyle{aa}


\longtab[1]{
	\begin{longtable}{ccc}

\caption{ \label{refparam} The reference values of the HARPS spectra used for the machine learning. T$_{\mathrm{eff}}$ and [Fe/H] have been derived photometrically by \citet{casag08} and \citet{neves12} respectively.} \\
\hline\hline
Star & [Fe/H] & T$_{\mathrm{eff}}$  \\
 & [dex] & [K] \\
 \hline
\endfirsthead
\caption{continued}\\
\hline\hline
Star & [Fe/H] & T$_{\mathrm{eff}}$  \\
 & [dex] & [K] \\
\hline
\endhead

\hline
\endfoot

Gl1 & -0.40 & 3528 \\ 
Gl54.1 & -0.40 & 2901 \\ 
Gl87 & -0.30 & 3565 \\ 
Gl105B & -0.14 & 3054 \\ 
HIP12961 & -0.12 & 3904 \\ 
LP771-95A & -0.51 & 3393 \\ 
GJ163 & 0.00 & 3223 \\ 
Gl176 & 0.02 & 3369 \\ 
GJ179 & 0.14 & 3076 \\ 
Gl191 & -0.79 & 3679 \\ 
Gl205 & 0.17 & 3497 \\ 
Gl213 & -0.19 & 3026 \\ 
Gl229 & -0.04 & 3586 \\ 
HIP31293 & -0.04 & 3312 \\ 
HIP31292 & -0.11 & 3158 \\ 
Gl250B & -0.09 & 3369 \\ 
Gl273 & -0.05 & 3107 \\ 
Gl300 & 0.09 & 2965 \\ 
GJ2066 & -0.09 & 3388 \\ 
GJ317 & 0.22 & 3130 \\
Gl341 & -0.14 & 3633 \\ 
GJ1125 & -0.15 & 3162 \\ 
Gl357 & -0.30 & 3335 \\ 
Gl358 & 0.01 & 3240 \\ 
Gl367 & -0.09 & 3452 \\ 
Gl382 & 0.04 & 3429 \\ 
Gl393 & -0.13 & 3396 \\ 
GJ3634 & -0.02 & 3332 \\ 
Gl413.1 & -0.06 & 3373 \\ 
Gl433 & -0.13 & 3450 \\ 
Gl436 & 0.01 & 3277 \\ 
Gl438 & -0.31 & 3536 \\ 
Gl447 & -0.23 & 2952 \\
Gl465 & -0.54 & 3382 \\ 
Gl479 & 0.05 & 3238 \\ 
Gl514 & -0.13 & 3574 \\ 
Gl526 & -0.18 & 3545 \\ 
Gl536 & -0.13 & 3546 \\ 
Gl555 & 0.13 & 2987 \\ 
Gl569A & 0.16 & 3235 \\ 
Gl581 & -0.18 & 3203 \\ 
Gl588 & 0.07 & 3284 \\ 
Gl618A & -0.05 & 3242 \\ 
Gl628 & -0.05 & 3107 \\ 
GJ1214 & 0.03 & 2856 \\ 
Gl667C & -0.47 & 3431 \\ 
Gl674 & -0.18 & 3284 \\ 
GJ676A & 0.10 & 3734 \\ 
Gl678.1A & -0.10 & 3611 \\ 
Gl680 & -0.07 & 3395 \\ 
Gl682 & 0.09 & 3002 \\ 
Gl686 & -0.29 & 3542 \\ 
Gl693 & -0.28 & 3188 \\ 
Gl699 & -0.59 & 3094 \\ 
Gl701 & -0.20 & 3535 \\ 
Gl752A & 0.04 & 3336 \\ 
Gl832 & -0.17 & 3450 \\ 
Gl846 & -0.08 & 3682 \\ 
Gl849 & 0.24 & 3200 \\ 
Gl876 & 0.14 & 3059 \\ 
Gl877 & -0.01 & 3266 \\ 
Gl880 & 0.05 & 3488 \\ 
Gl887 & -0.20 & 3560 \\ 
Gl908 & -0.38 & 3587 \\ 
LTT9759 & 0.17 & 3316 \\

\end{longtable}
}

\clearpage

\begin{appendix} 
\section{Evaluation of our pseudo-EW measurements}
\label{A}
 
\begin{figure}
\centering
$\begin{array}{c}
\includegraphics[width=8cm]{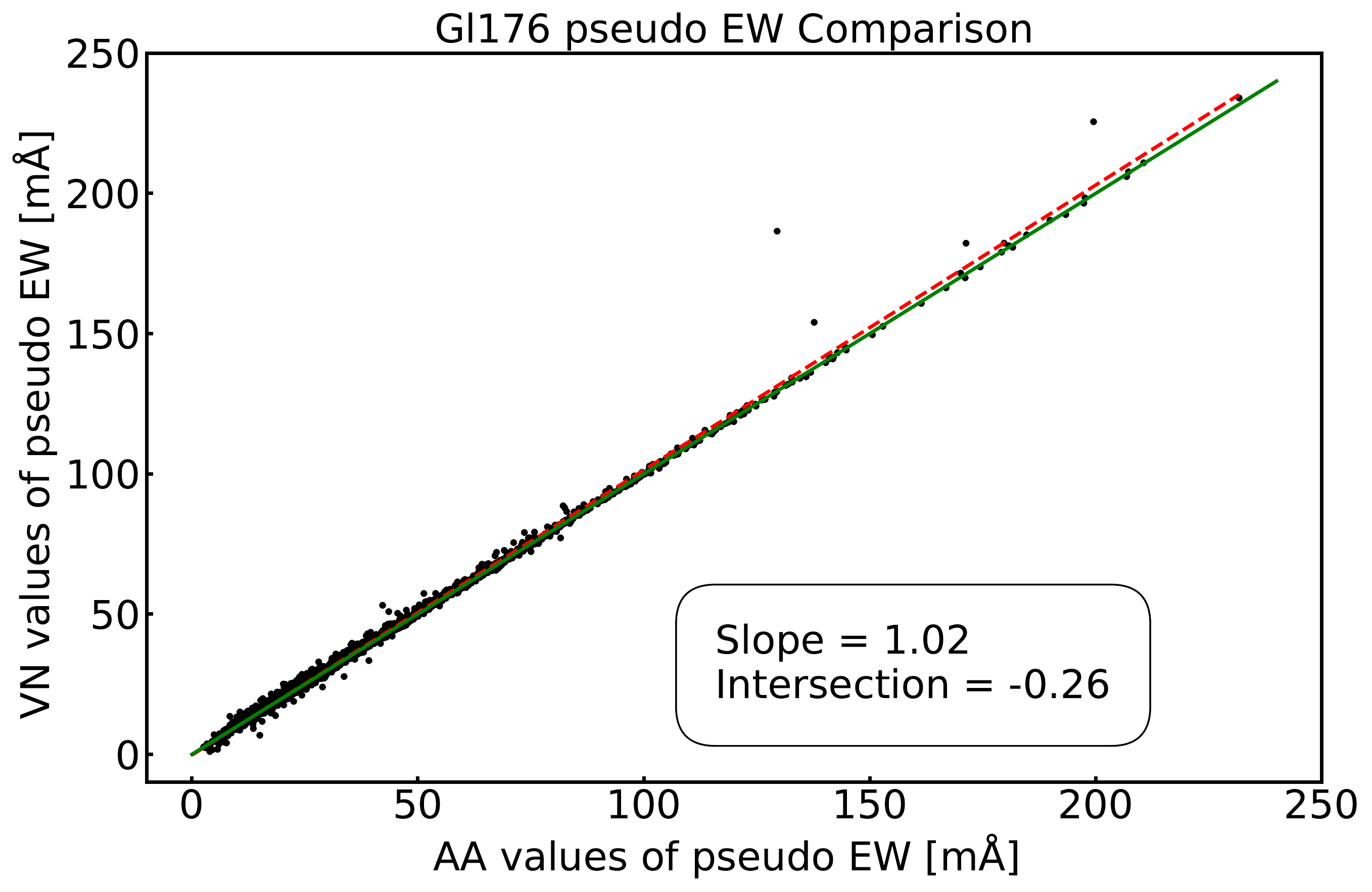} \\
\includegraphics[width=8cm]{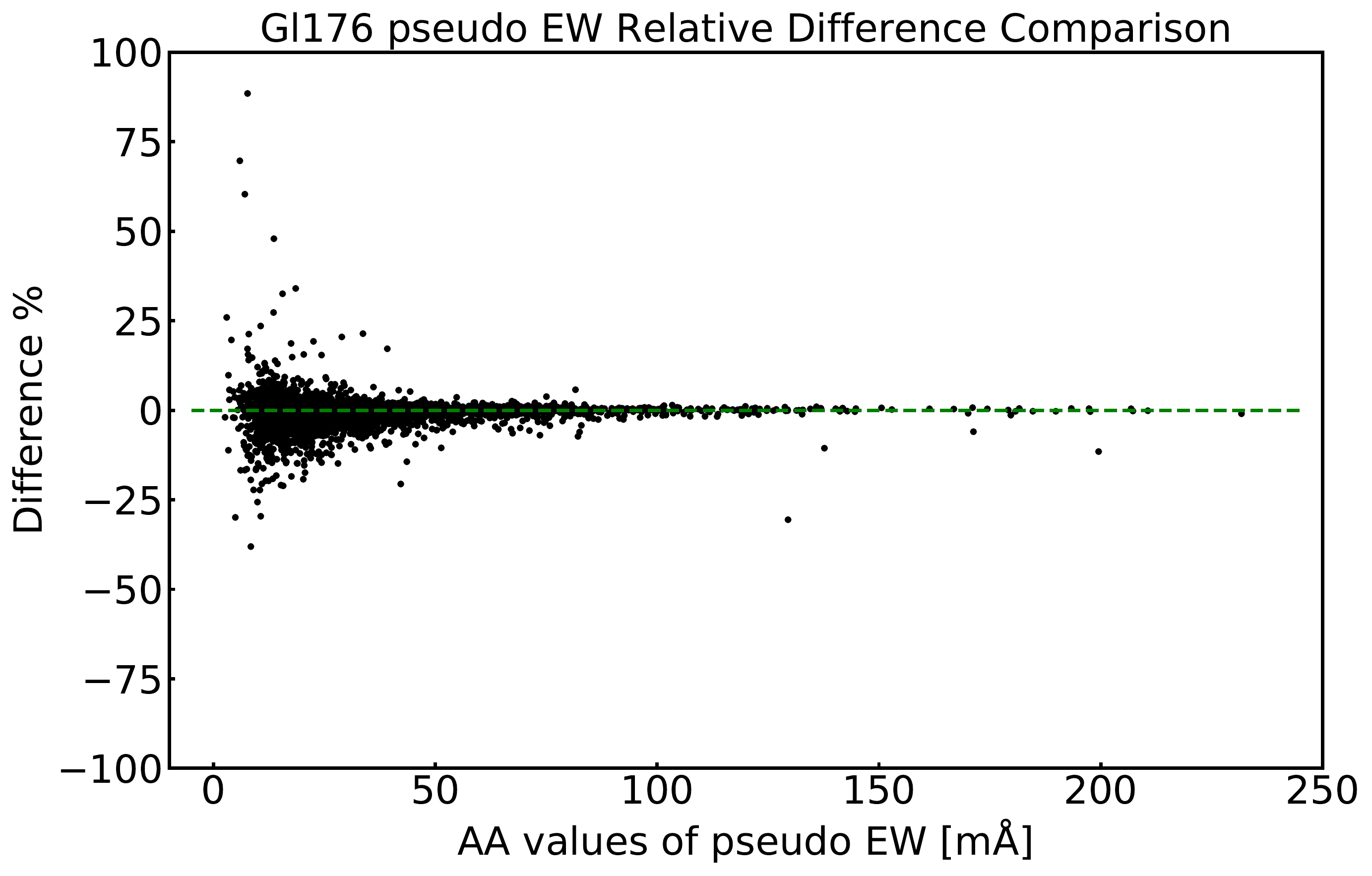} \\
\end{array}$
\caption{Upper panel: Comparison between the pseudo-EW values measured by our tool and MCAL code by \citet{neves14}. Lower panel: The percentage difference of the pseudo-EW values plotted against our values.}
\label{comp} 
\end{figure}

\begin{figure}
\centering
$\begin{array}{c}
\includegraphics[width=8cm]{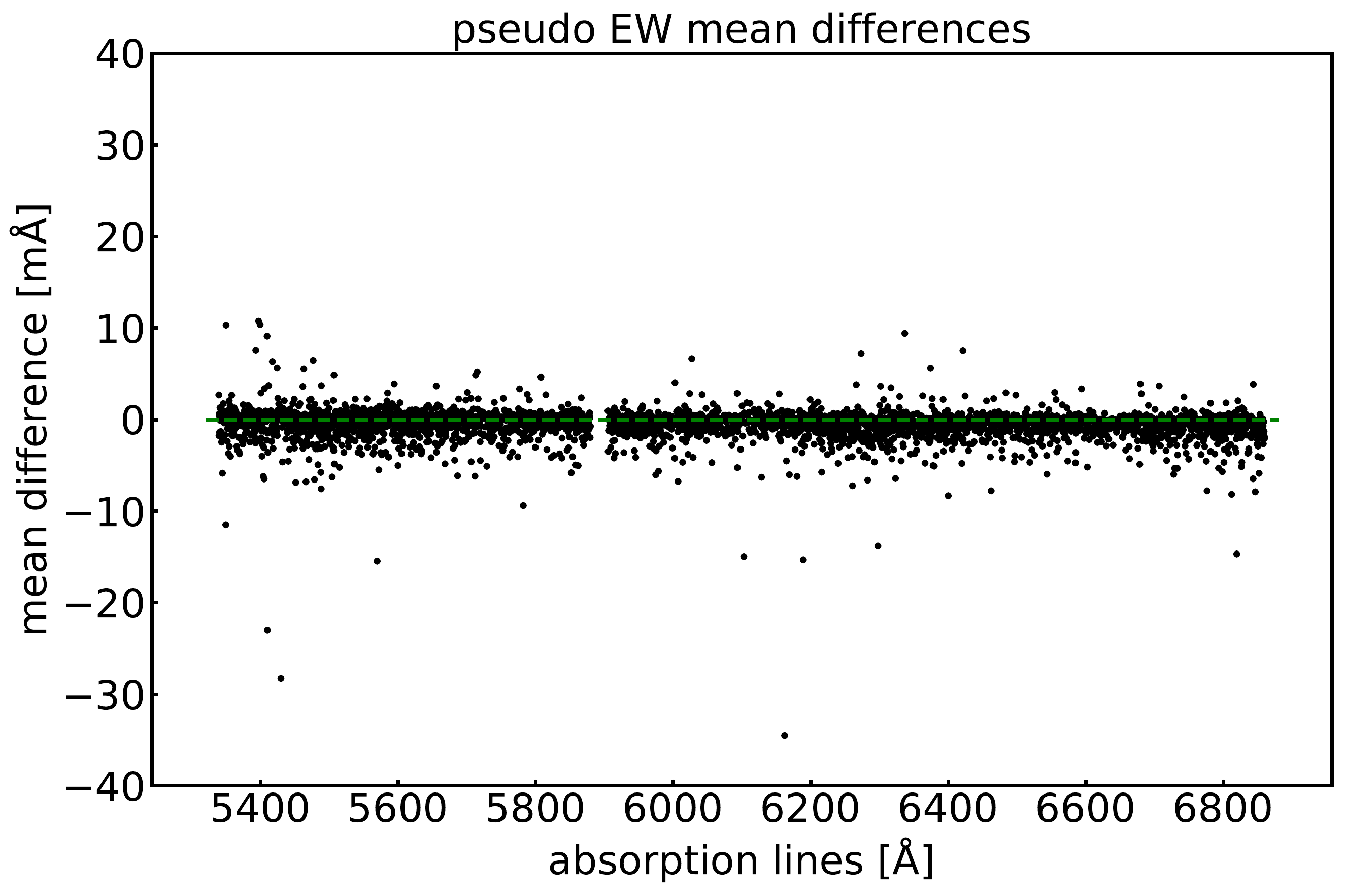} \\
\includegraphics[width=8cm]{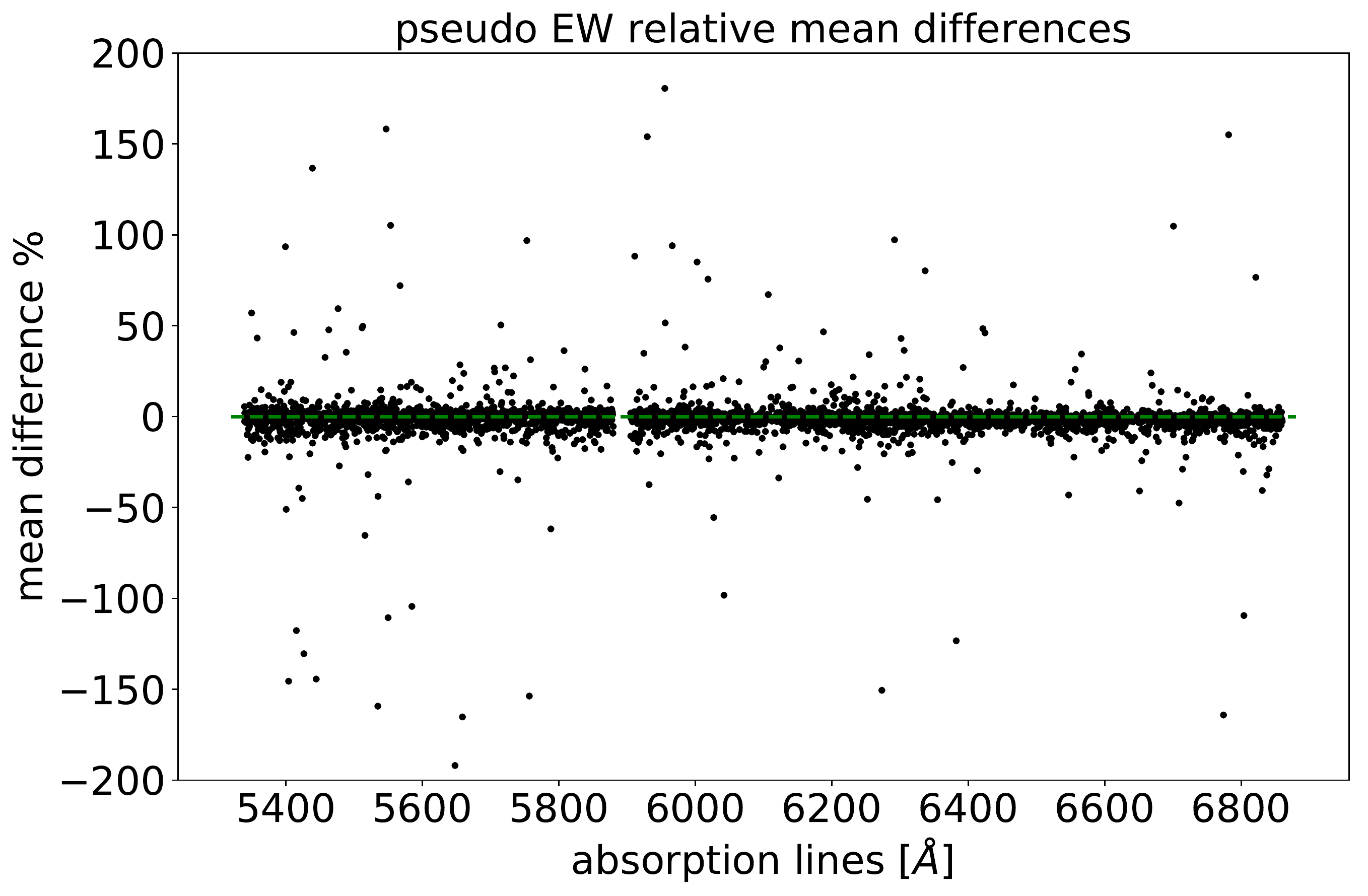} \\
\end{array}$
\caption{Upper panel: The difference of each line as averaged by the measurements of all stars. Lower panel: The percentage difference of each line as averaged by the measurements of all stars.}
\label{d} 
\end{figure}

We calculated the pseudo EWs of 4104 lines for the 110 stars of the total HARPS sample. Here we compare our values with the ones obtained by MCAL code. In the upper panel of Fig.~\ref{comp}, we present the comparison of all the pseudo-EW values for the star Gl176 as an example. The units of pseudo EWs are m$\AA$. Inside the plots, AA stands for our measurements and VN stands for the measurements by \citet{neves14}. The slope in the diagrams of most stars is almost identical with the identity line, with only few pseudo EWs having considerably different values. 
In the lower panel of Fig.~\ref{comp} we show the relative difference (the percentage) of the values against our values. A scatter appears for the pseudo-EW values smaller than 30 m$\AA$, which is normal, as the relative difference for these narrow lines is greater. In contrast, nearly all lines broader than 50 m$\AA$ are measured with significant agreement. 
The actual quality test for measuring the pseudo EWs, comes from the following comparison. We plot the mean differences and mean relative differences between our method and the code by \citet{neves14} for each line averaged by all the stars, to see how all the lines are measured. 
The result is very good as it can be seen in Fig.~\ref{d}, with only 184 lines out of 4104 showing a mean relative difference greater than $\pm$15\%.

\end{appendix}

\end{document}